\documentclass[aps,prapplied,reprint,longbibliography,amsmath,amssymb]{revtex4-2}

\usepackage{graphicx}
\usepackage{bm}
\usepackage{color}
\usepackage{hyperref}

\begin{document}

\title{Hot-Carrier Distribution Spectroscopy by Transconductance
       in Two-Dimensional Field-Effect Transistors}

\author{Katsunori Wakabayashi}
\affiliation{Research Center for Materials Nanoarchitectonics (MANA), National Institute for Materials Science (NIMS), Namiki 1-1, Tsukuba 305-0044, Japan}

\date{\today}

\begin{abstract}
The transconductance $g_m = dI_D/dV_G$ of a field-effect transistor (FET)
is conventionally read as a proxy for carrier density.
We show that it is instead a spectroscopic probe of the carrier distribution:
because $g_m$ weights the spectral current $j(E)$ by the gate-voltage
derivative $\partial f(E)/\partial V_G$ and integrates over energy, it is
sensitive to the \emph{shape} of $f(E)$, not merely its integrated weight $n$.
We develop an energy-resolved transport framework for two-dimensional (2D)
FETs and, within a gate-independent spectral-kernel approximation, derive the
decomposition $g_m = g_m^{(n)} + g_m^{(\alpha)}$ into the conventional
density-modulation term $g_m^{(n)}$ and a distribution-shape-driven term
$g_m^{(\alpha)}$.
The latter, obtained as the residual after subtracting the smooth
density-modulation background from the measured $g_m$, exhibits a
characteristic anomalous peak at a gate voltage $V_G^{\rm pk}$.
This peak has no counterpart in equilibrium transport and
\emph{cannot be explained by carrier density modulation alone}.
With the spectral kernel calibrated, the peak position and
height---extracted from standard DC/lock-in $g_m$ sweeps---constrain the
hot-carrier energy $E_0$, spectral width $\sigma$, and generation threshold
$n_c$, realizing a steady-state, all-electrical spectroscopy of the carrier
distribution.
An optional time-resolved extension further recovers the carrier relaxation
time $\tau$ from the transient response following a pump excitation,
establishing the 2D FET as a distribution-function spectrometer that
requires no optical readout.
\end{abstract}

\maketitle

\section{Introduction}

Two-dimensional (2D) semiconductors such as transition metal
dichalcogenides (TMDs)~\cite{Mak2010,Wang2012,Chhowalla2013}
and their van der Waals heterostructures have emerged as a rich arena for
exploring electronic transport in reduced dimensions~\cite{Radisavljevic2011,Kang2015}.
Field-effect transistors (FETs) fabricated from these materials
simultaneously serve as testbeds for fundamental physics and as building
blocks of next-generation devices~\cite{Liao2010,Li2014,Fiori2014,Li2024,Chen2024}.

In the standard description of FET transport, the drain current is written as
\begin{equation}
  I_D = \frac{qW}{L}\,n\,\mu\,V_D,
  \label{eq:standard}
\end{equation}
where $q$ is the elementary charge, $W$ and $L$ are the channel width and
length, $n$ is the areal carrier density, $\mu$ is the mobility, and $V_D$
is the source--drain voltage~\cite{Sze2007}.
This relation implicitly assumes that carriers obey an equilibrium or
quasi-equilibrium distribution, so that $\mu$ is a single scalar that
summarizes all microscopic scattering.
Under strong source--drain fields, carrier injection, or optical pumping,
however, the distribution function acquires a hot-carrier tail whose
energy is well above the lattice temperature.
In such nonequilibrium states, the shape of the distribution---not merely
its integrated weight $n$---governs transport, rendering
Eq.~(\ref{eq:standard}) inadequate~\cite{Lundstrom2000,Datta1995}.

Nonequilibrium carriers in 2D materials have been studied extensively
by ultrafast optical spectroscopy~\cite{Massicotte2021}.
Pump--probe experiments on graphene~\cite{Sun2008} and
MoS$_2$~\cite{Pogna2016,Shi2014,Wang2021,Lee2021}
reveal hot-carrier relaxation times in the range $\tau \sim 0.1$--$10$~ps,
reflecting carrier--phonon scattering dynamics.
Yet such optical probes are rarely combined with operando electrical
gating, and they do not directly access the transport-relevant,
velocity-weighted distribution that governs $I_D$.

Here we develop an \emph{energy-resolved} transport theory for 2D FETs
that connects the shape of $f(E)$ to measurable device characteristics.
Our starting point is a spectroscopic reformulation of transport: within an
energy-resolved transport description with a gate-independent spectral
current kernel $j(E)$, the transconductance takes the form
\begin{equation}
  g_m = \frac{qW}{L}\int_0^\infty j(E)\,\frac{\partial f(E;V_G)}{\partial V_G}\,dE,
  \label{eq:gm_spectral_intro}
\end{equation}
which makes explicit that \textit{the transconductance does not merely reflect
carrier density modulation but probes the energy distribution of carriers}.
Equation~(\ref{eq:gm_spectral_intro}) is itself a natural consequence of
writing the current as a spectral integral and differentiating with respect
to $V_G$.
It is a close relative of the Landauer--B\"uttiker
formula~\cite{Datta1995,Buttiker1986} and of the transport distribution
function of thermoelectric transport~\cite{MahanSofo1996}.
More broadly, it instantiates a general principle---that a derivative of the
conductance resolves spectral structure---which underlies
scanning-tunneling~\cite{TersoffHamann1985} and quantum-dot
transconductance~\cite{Kouwenhoven2001} spectroscopies as well as the
electrical measurement of the quasiparticle distribution in mesoscopic
wires~\cite{Pothier1997}.

\begin{figure*}[t]
  \centering
  \includegraphics[width=\textwidth]{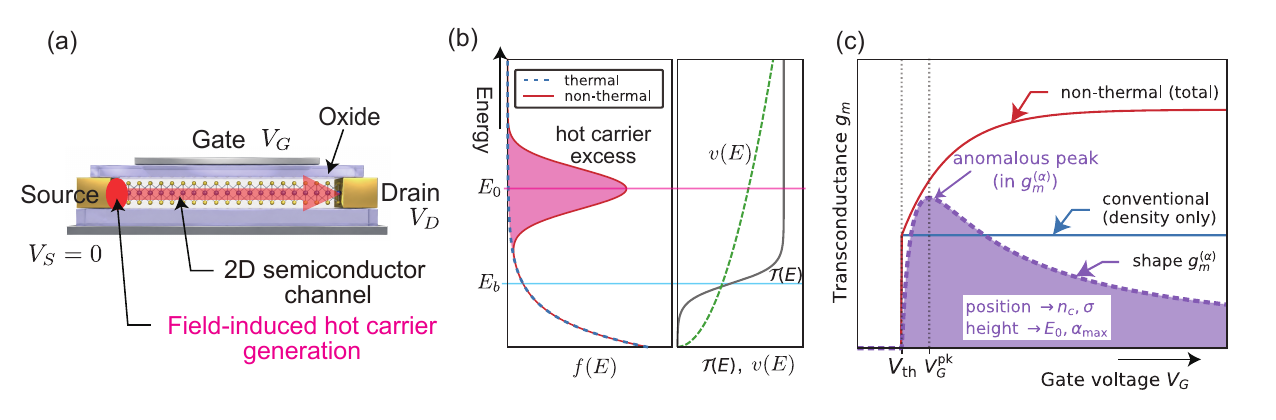}
  \caption{%
    Concept, device, and measurement principle (schematic).
    (a) A gated monolayer 2D-semiconductor field-effect transistor: a gate
    voltage tunes the carrier density in the channel between source and drain
    contacts, while field-induced hot-carrier generation drives the distribution out of
    equilibrium.
    (b) Physical mechanism (energy on the vertical axis).
    \emph{Left:} the carrier distribution $f(E)$ is either thermal (dotted) or
    non-thermal (red), the latter carrying a localized hot-carrier excess near
    $E_0$.
    \emph{Right:} the transport kernel---an energy-selective transmission
    $\mathcal{T}(E)$ that turns on near the onset $E_b$ and a velocity $v(E)$
    that grows with energy---weights high-energy carriers more heavily, so each
    carrier in the excess near $E_0 > E_b$ contributes more current.
    (c) Measurement principle (schematic): a conventional, density-only response
    is featureless in $V_G$ (blue), and the total non-thermal $g_m$ (red) rises
    monotonically. The non-thermal signature is the shape-driven contribution
    $g_m^{(\alpha)}$ (purple, shaded), which produces an anomalous peak whose
    position and height read out the spectral parameters of
    $f(E)$---transconductance spectroscopy.
    (In the representative case shown here the total measured $g_m$ itself
    remains monotone; the peak is isolated in $g_m^{(\alpha)}$, as quantified in
    Fig.~\ref{fig:steadystate}.)
    The response turns on at the threshold voltage $V_{\rm th}$
    [Eq.~(\ref{eq:n_gate})], where the channel begins to populate, and the
    anomaly peaks at $V_G^{\rm pk}$ [Eq.~(\ref{eq:VGpk})].
  }
  \label{fig:concept}
\end{figure*}

Our contribution is to apply this principle to the \emph{shape of a
nonequilibrium distribution} in a 2D FET, where it yields the specific,
measurable anomaly developed below.
In a related line of work, the same transconductance-as-probe principle is
applied to the valley degree of freedom in WSe$_2$, using $g_m$ to read out
valley thermodynamics in multilayer films~\cite{Wakabayashi2026b} and to guide
valley engineering in bilayer gate-all-around transistors~\cite{Wakabayashi2026a}.
A dynamical extension of the same idea uses the transconductance to extract the
intervalley relaxation time of WSe$_2$ transistors~\cite{Wakabayashi2026c}.
Corrections from a $V_G$-dependent kernel $j(E,V_G)$ (e.g.\ barrier
lowering, contact modulation, or band bending) are additive and
discussed in Sec.~\ref{sec:discussion}.

For a Boltzmann distribution, $\partial f_{\rm eq}/\partial V_G$ is featureless
in energy, and $g_m$ is flat---the conventional result.
For a nonequilibrium distribution with a hot-carrier peak at $E_0$,
$\partial f_{\rm neq}/\partial V_G$ acquires structure at $E_0$
and $g_m$ reports it as a measurable anomaly in the shape-resolved residual.
The Gaussian hot-carrier peak adopted below is only a minimal model: the
mechanism applies to any localized nonthermal excess $\delta f(E)$ above the
transport onset, not to a specific line shape.
Figure~\ref{fig:concept}(a) shows the device geometry, (b) the spectral
mechanism, and (c) the resulting measurement principle.

Our main results are threefold.
(i) The drain current takes the compact form
$I_D = (qW/L)\,n\,\bar{v}(\alpha)$, with an effective velocity $\bar{v}(\alpha)$
set by a single parameter $\alpha$ that quantifies the hot-carrier population.
(ii) Within the same spectral-kernel approximation, $g_m$ admits a
transparent decomposition $g_m = g_m^{(n)} + g_m^{(\alpha)}$.
The new term $g_m^{(\alpha)}$ produces an anomalous peak in the
shape-resolved residual---obtained after calibrating and subtracting the
smooth density-modulation background---that \emph{cannot be explained by
carrier density modulation} and directly constrains the spectral parameters
$\{E_0,\sigma,n_c\}$.
This peak carries a decisive drain-voltage fingerprint---its position shifts
to lower $V_G$ and its height saturates with $V_D$---that no separable,
quasi-equilibrium mobility-rolloff model can reproduce.
(iii) A time-resolved electrical measurement after a pump excitation
yields the carrier relaxation time $\tau$, completing an all-electrical
spectroscopy of the nonequilibrium distribution.

The remainder of this paper is organized as follows.
Section~\ref{sec:model} develops the energy-resolved transport framework and
derives the decomposition $g_m = g_m^{(n)} + g_m^{(\alpha)}$.
Section~\ref{sec:results} presents the anomalous $g_m^{(\alpha)}$ peak, its
drain-voltage fingerprint, and the resulting extraction of the spectral
parameters $\{E_0,\sigma,n_c\}$.
Section~\ref{sec:discussion} places the framework in the context of prior
hot-carrier work and addresses corrections from a gate-dependent kernel, and
Sec.~\ref{sec:conclusion} concludes.
Appendices~\ref{app:timedomain}--\ref{app:alphaforms} present the
time-resolved extension, a fit to synthetic data, and the robustness of the
anomaly against modeling choices.

\section{Model and Theory}
\label{sec:model}

\subsection{Energy-resolved drain current}

For definiteness, we formulate the model for an n-type 2D FET, where $E \geq 0$
denotes the electron kinetic energy measured from the conduction-band minimum.
The same formalism applies to p-type devices by interpreting $E$ as the hole
kinetic energy measured from the valence-band maximum.

For a 2D semiconductor of width $W$ and length $L$, we write the drain
current as an integral over carrier energy~\cite{Datta1995,Lundstrom2000}:
\begin{equation}
  I_D = \frac{qW}{L}\int_0^\infty D(E)\,v(E)\,\mathcal{T}(E)\,f(E)\,dE,
  \label{eq:ID_general}
\end{equation}
where $D(E)$ is the density of states, $v(E)$ is the group velocity,
$\mathcal{T}(E)$ is an energy-selective transmission probability, and $f(E)$
is the carrier distribution function.
For a 2D system with parabolic dispersion, $D(E) = D_0 = g\,m^*/(\pi\hbar^2)$
is constant (with $g$ the spin--valley degeneracy and $m^*$ the effective
mass), and the group velocity is
\begin{equation}
  v(E) = v_T\sqrt{E/k_BT},
  \label{eq:velocity}
\end{equation}
where $v_T = \sqrt{k_BT/m^*}$ is a reference (thermal) velocity scale.%
\footnote{The exact group velocity of the 2D parabolic band is
$v(E) = \sqrt{2E/m^*}$; here we absorb the $O(1)$ numerical prefactor
(the $\sqrt{2}$ and the angular average of the transport-direction component)
into the definition of $v_T$, since only the $\sqrt{E}$ energy dependence
enters the spectral shape---overall constants cancel in the velocity ratios
$\bar{v}(\alpha)$ and in $g_m$.}
The spectral current $j(E) \equiv v(E)\mathcal{T}(E)$ acts as an energy
filter: it weights high-energy carriers more heavily in proportion
to their velocity.

We model the energy selectivity of the transmission by a smooth onset,
\begin{equation}
  \mathcal{T}(E) = \left[1 + \exp\!\left(\frac{E_b - E}{\Delta}\right)\right]^{-1},
  \label{eq:T}
\end{equation}
a smooth step centered at an effective onset energy $E_b$ with a broadening
parameter $\Delta$.
We do \emph{not} identify $E_b$ with a unique microscopic barrier height.
Rather, $E_b$ is an effective crossover energy of the transport kernel that
folds contact injection (e.g.\ a Schottky barrier), electrostatic bottlenecks
(channel pinch-off, i.e.\ the saddle point of the channel potential), and
energy-dependent scattering into a single energy-filtering scale.
Carriers with $E < E_b$ contribute weakly to the drain current, while those
with $E > E_b$ are efficiently transmitted~\cite{Lundstrom2000}.
Both $E_b$ and $\Delta$ are thus effective transport-kernel parameters, to be
calibrated from the equilibrium response $\bar{v}_{\rm eq}(T)$ rather than
identified with an independent microscopic barrier.
Importantly, none of our central results depend on this specific
functional form. They require only that the spectral current
$j(E) = v(E)\mathcal{T}(E)$ be a smooth, monotonically increasing filter
that favors high-energy carriers, so that a hot-carrier population at
$E_0 > E_b$ is preferentially transmitted.
Any transmission with these properties yields the same anomalous
$g_m^{(\alpha)}$ peak, in the same spirit as the robustness against the choice
of $\alpha(n)$ demonstrated in Appendix~\ref{app:alphaforms}.

\subsection{Carrier distribution functions}

We compare two classes of distribution, both normalized to the same
areal carrier density $n$ via $\int_0^\infty f(E)\,dE = n$.
Here $f(E)$ is a carrier spectral density (carriers per unit area and energy),
not a dimensionless occupation probability; accordingly it carries dimensions
of $[n]/[\text{energy}]$, with the density of states $D_0$ absorbed into the
current prefactor as noted below.

\paragraph{Equilibrium (Boltzmann).}
\begin{equation}
  f_{\rm eq}(E) = A_{\rm eq}\,e^{-E/k_BT},
  \label{eq:feq}
\end{equation}
where $A_{\rm eq} = n/N_{\rm eq}$ and $N_{\rm eq} = \int_0^\infty e^{-E/k_BT}dE = k_BT$.
We adopt the Boltzmann (nondegenerate) form for analytical transparency.
It is the leading approximation to the Fermi--Dirac distribution
$f_{\rm FD}(E) = \{1+\exp[(E-E_F)/k_BT]\}^{-1}$ when the Fermi level lies
below the transport states by more than a few $k_BT$, i.e.\ in the moderately
doped, nondegenerate regime relevant to the density scale
$n_c\sim10^{12}$~cm$^{-2}$ considered here.
For strongly degenerate channels the Boltzmann factor $e^{-E/k_BT}$ in
Eqs.~(\ref{eq:feq})--(\ref{eq:fneq}) should be replaced by $f_{\rm FD}(E)$.
This changes only the numerical values of the spectral moments $N_{\rm eq}$
and $J_{\rm eq}$, leaving the spectroscopic identity
Eq.~(\ref{eq:gm_spectral}) and the decomposition
$g_m = g_m^{(n)} + g_m^{(\alpha)}$ unchanged in form.
The hot-carrier component, centered well above $E_F$ at $E_0 \gg k_BT$, is
nondegenerate in either description.

\paragraph{Nonequilibrium (Boltzmann + hot-carrier Gaussian).}
\begin{equation}
  f_{\rm neq}(E) = A_{\rm neq}\!\left[
    e^{-E/k_BT} + \alpha\,e^{-(E-E_0)^2/\sigma^2}
  \right],
  \label{eq:fneq}
\end{equation}
where $A_{\rm neq} = n/(N_{\rm eq} + \alpha N_{\rm neq})$,
$N_{\rm neq} = \int_0^\infty e^{-(E-E_0)^2/\sigma^2}dE \approx \sigma\sqrt{\pi}$
(for $E_0 \gg \sigma$), and $\alpha \geq 0$ controls the amplitude of the
Gaussian hot-carrier bump centered at energy $E_0 > E_b$ with width $\sigma$.
The constraint $\int f(E)\,dE = n$---where the constant $D_0$ is absorbed into
the current prefactor, so that it cancels in $\bar{v}$ and in every $g_m$ ratio
below---is satisfied by construction via $A_{\rm neq}$.
The Gaussian is a minimal representation of a localized nonthermal excess at
$E_0$.

Physically, such a peaked nonthermal population can be established in several
ways: high-field acceleration in the drain-side region of a short-channel FET
followed by energy-selective optical-phonon or intervalley scattering; carrier
injection over a contact or barrier bottleneck that admits a narrow energy
window; impact-ionization-like generation near a characteristic energy; or
optical pumping at a fixed photon energy.
In each case carriers accumulate around a characteristic energy $E_0$ set by
the dominant generation and relaxation bottleneck rather than by the lattice
temperature.
We do not commit to a specific microscopic mechanism; the Gaussian models only
the resulting localized excess.
The spectroscopic mechanism developed below relies only on the presence
of such an excess well above the transport onset $E_b$.
The same $g_m^{(\alpha)}$ anomaly therefore arises for any nonthermal excess
$\delta f(E)$ that is sharply peaked in this energy range, irrespective of its
detailed shape.

The physical meaning of $\alpha$ is the relative number of hot carriers
in the Gaussian component: at $\alpha = 0$ the distribution is purely
thermal, while increasing $\alpha$ builds up a high-energy tail
at $E_0 \gg k_BT$.

\subsection{Effective drift velocity and current decomposition}

Substituting Eqs.~(\ref{eq:feq})--(\ref{eq:fneq}) into Eq.~(\ref{eq:ID_general})
and using the linearity of the integral, we define spectral current moments
\begin{align}
  J_{\rm eq}  &= \int_0^\infty j(E)\,e^{-E/k_BT}\,dE, \label{eq:Jeq} \\
  J_{\rm neq} &= \int_0^\infty j(E)\,e^{-(E-E_0)^2/\sigma^2}\,dE. \label{eq:Jneq}
\end{align}
The drain current then takes the compact form
\begin{equation}
  I_D = \frac{qW}{L}\,n\,\bar{v}(\alpha),
  \label{eq:ID_decomposed}
\end{equation}
where the \emph{effective drift velocity} is
\begin{equation}
  \bar{v}(\alpha) = \frac{J_{\rm eq} + \alpha J_{\rm neq}}{N_{\rm eq} + \alpha N_{\rm neq}}.
  \label{eq:vbar}
\end{equation}
Equivalently, $\bar{v}(\alpha) = \int_0^\infty j(E)f(E)\,dE \big/
\int_0^\infty f(E)\,dE$ is simply the carrier-weighted average of the spectral
current $j(E)$: the normalization $A_{\rm neq}$ and the density $n$ cancel
between numerator and denominator, leaving each moment split into a thermal and
an $\alpha$-weighted hot-carrier part.
Equation~(\ref{eq:ID_decomposed}) preserves the form of the conventional
relation $I_D \propto nv_{\rm drift}$, but now $\bar{v}(\alpha)$ is
distribution-shape-sensitive: it interpolates between the pure-Boltzmann
limit $\bar{v}(0) = J_{\rm eq}/N_{\rm eq} \equiv \bar{v}_{\rm eq}$ and the
Gaussian limit $\bar{v}(\infty) = J_{\rm neq}/N_{\rm neq} \equiv \bar{v}_{\rm neq}$.
Because hot carriers at $E_0 \gg E_b$ are above the barrier ($\mathcal{T}(E_0)\approx 1$)
and have higher velocity ($v(E_0) \propto \sqrt{E_0}$), we have
$\bar{v}_{\rm neq} \gg \bar{v}_{\rm eq}$, so even a small $\alpha$
significantly boosts $I_D$.

\begin{figure}[tbp]
  \centering
  \includegraphics[width=\columnwidth]{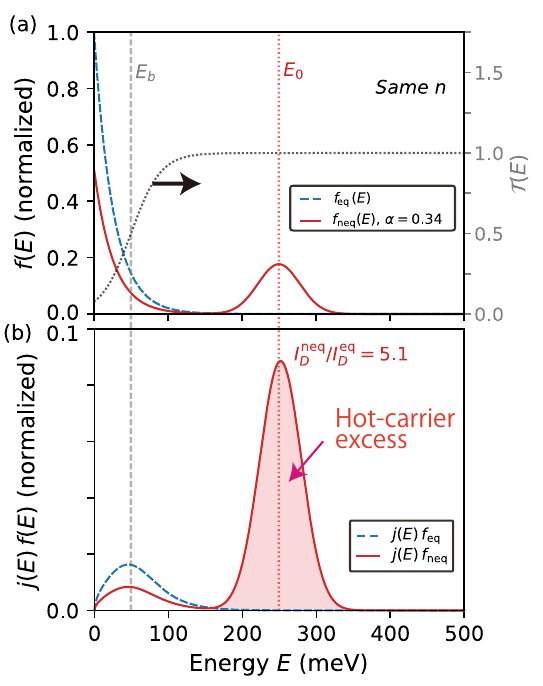}
  \caption{%
    (a) Carrier distribution functions $f(E)$ for the equilibrium (Boltzmann,
    blue dashed) and nonequilibrium (Boltzmann + Gaussian bump, red solid,
    $\alpha = 0.34$) cases, both normalized to the same carrier density $n$.
    The gray dotted curve on the right axis shows the transmission
    $\mathcal{T}(E)$; vertical dashed lines mark the onset energy $E_b$ and the
    hot-carrier peak energy $E_0$.
    (b) Spectral current integrand $j(E)f(E)$ for both distributions (shared
    energy axis with (a)).
    The shaded region is the hot-carrier excess contribution to the current.
    Despite identical carrier densities, the nonequilibrium current is enhanced
    by $I_D^{\rm neq}/I_D^{\rm eq}\approx 5.1$ at this density, reflecting the
    larger per-carrier current of the hot-carrier excess near $E_0$
    ($\bar{v}_{\rm neq}/\bar{v}_{\rm eq}\approx 9.4$ for the pure Gaussian
    component).
  }
  \label{fig:distributions}
\end{figure}

\subsection{Gate-controlled carrier density and nonequilibrium parameter}

Above threshold the carrier density follows the linear gate relation,
\begin{equation}
  n(V_G) = \frac{C_{\rm ox}}{q}(V_G - V_{\rm th}),
  \label{eq:n_gate}
\end{equation}
where $C_{\rm ox}$ is the oxide capacitance per unit area and $V_{\rm th}$
is the threshold voltage.
This is an idealized sharp threshold ($n\equiv0$ below $V_{\rm th}$, linear
above), so $dn/dV_G$ jumps discontinuously at $V_{\rm th}$.
Through the decomposition derived below, this produces two turn-on features in
$g_m$ exactly at $V_{\rm th}$: a step in the density term
$g_m^{(n)}\propto dn/dV_G$ (which jumps from zero to a constant), and---in the
non-thermal case---a kink in the shape term $g_m^{(\alpha)}\propto n$ (whose
slope breaks where $n$ begins to rise from zero).
Both are artifacts of the idealized threshold: in a real device, subthreshold
conduction ($n\sim e^{qV_G/k_BT}$ below $V_{\rm th}$) smooths them into a
gradual turn-on.
In any case, these features sit at $V_{\rm th}$ and are well separated from---and
do not affect---the anomalous peak at $V_G^{\rm pk}$.

The hot-carrier amplitude $\alpha$ depends on both $V_G$ and $V_D$.
The key qualitative property required for an anomalous peak in $g_m^{(\alpha)}$ is
that $\alpha(n)$ \emph{rises from zero at small $n$ and saturates as
$n\to\infty$}.
A broad class of such monotone-and-saturating functions generically
produces a localized peak in the distribution-shape contribution
$g_m^{(\alpha)}$ derived below, because $d\alpha/dn$ inherits the bell
shape of the derivative of a saturating function.
The qualitative peak position, height, and $V_D$-dependence are robust
fingerprints of this mechanism, only weakly dependent on the precise
functional form of $\alpha(n)$.
Alternative analytic forms (hyperbolic-tangent, exponential saturation,
and a Hill function in $n$) yield quantitatively similar $g_m^{(\alpha)}(V_G)$
peaks, differing only in the prefactor of the peak density, as verified
explicitly in Appendix~\ref{app:alphaforms} (Fig.~\ref{fig:robustness}).

For analytical tractability we adopt the representative phenomenological
model
\begin{equation}
  \alpha(V_G,V_D) = \alpha_{\rm max}\,
    \frac{V_D^2}{V_D^2 + V_c^2}\,
    \frac{n(V_G)}{n(V_G) + n_c},
  \label{eq:alpha}
\end{equation}
which is proportional to the power density (through $V_D$) and to the
density of impact-ionizing or inelastically scattered carriers (through
$n$).
Here $\alpha_{\rm max}$ is the saturation amplitude, $V_c$ is the
drain-voltage crossover, and $n_c$ is the carrier-density crossover for
hot-carrier generation.
The high-field crossover at $V_c$ parallels the onset of velocity saturation
observed in monolayer MoS$_2$~\cite{Smithe2018,Wang2025}.
In real devices $n_c$ is not a universal material constant but an effective
crossover density set by the channel length, contact field
concentration~\cite{Allain2015}, dielectric environment, heat dissipation, and
the dominant inelastic-scattering pathway~\cite{Kaasbjerg2012}.
The value adopted below should therefore be viewed as a representative
effective scale for strongly gated TMD FETs.
For $n \ll n_c$, $\alpha \propto n$ (few carriers, few collisions);
for $n \gg n_c$, $\alpha \to \alpha_{\rm max}h(V_D)$ (saturation).

We stress that Eq.~(\ref{eq:alpha}) is a transparent phenomenological
interpolation, not a microscopic derivation, and that the two factors are
chosen for their limiting behavior rather than for a specific functional form.
The drain-voltage factor $h(V_D) = V_D^2/(V_D^2+V_c^2)$ represents field-driven
hot-carrier generation that switches on with the power input at low bias and
saturates once the high-field region spans the channel. Any monotonic function
rising from zero and saturating at large $V_D$ yields the same qualitative
$g_m^{(\alpha)}(V_D)$.
The density factor $n/(n+n_c)$ expresses that at low density the hot-carrier
population grows in proportion to the number of available carriers (and hence
collisions), whereas at high density it saturates as phase-space restriction,
screening, and energy-relaxation bottlenecks limit further accumulation.
What matters for the anomalous peak is only that $\alpha(n)$ rise from zero and
saturate, so that $d\alpha/dn$ is bell-shaped. We verify explicitly in
Appendix~\ref{app:alphaforms} (Fig.~\ref{fig:robustness}) that
hyperbolic-tangent, exponential, and Hill forms all produce the same peak, with
only minor quantitative differences.
A first-principles kinetic treatment of $\alpha(n,V_D)$ from the coupled
hot-carrier generation--relaxation balance is beyond the scope of the present
work and is left for future study.

\subsection{Transconductance as a spectral probe}

Before decomposing $g_m$, we establish its fundamental spectroscopic content.
Differentiating Eq.~(\ref{eq:ID_general}) with respect to $V_G$ gives, by the
product rule, two contributions,
\begin{equation}
  g_m = \frac{qW}{L}\int_0^\infty\!\left[
    \frac{\partial j(E;V_G)}{\partial V_G}\,f(E;V_G)
    + j(E;V_G)\,\frac{\partial f(E;V_G)}{\partial V_G}
  \right]dE,
  \label{eq:gm_full}
\end{equation}
a \emph{kernel-modulation} term and a \emph{distribution-modulation} term.
Within the present transport description the spectral current kernel
$j(E) = v(E)\mathcal{T}(E)$ [Eqs.~(\ref{eq:velocity})--(\ref{eq:T})] has no
explicit gate dependence, so $\partial j/\partial V_G = 0$, the first term
vanishes, and we obtain
\begin{equation}
  g_m = \frac{qW}{L}\int_0^\infty j(E)\,\frac{\partial f(E;V_G)}{\partial V_G}\,dE.
  \label{eq:gm_spectral}
\end{equation}
The kernel-modulation term is nonzero when the gate lowers the barrier or
modulates the contacts [$\mathcal{T}=\mathcal{T}(E,V_G)$], induces band
bending, or otherwise renders $j$ gate-dependent.
This term is additive and is estimated in Sec.~\ref{sec:discussion}.
Equation~(\ref{eq:gm_spectral}) shows that the distribution-function
contribution to $g_m$ is exactly the spectral current $j(E)$ weighted by
$\partial f/\partial V_G$: a \emph{spectrally resolved} response function
of the distribution.
For $f_{\rm eq}$, the derivative $\partial f_{\rm eq}/\partial V_G \propto e^{-E/k_BT}$
is featureless, and $g_m$ depends only on the total carrier density.
For $f_{\rm neq}$, $\partial f_{\rm neq}/\partial V_G$ acquires additional
structure at $E_0$ whenever $\alpha$ depends on $V_G$ (i.e.\ on $n$),
and $g_m$ directly reports this spectral feature---even though no optical
measurement is performed.
This is the central spectroscopic principle of the paper.

Decomposing $g_m = dI_D/dV_G$ explicitly via
Eq.~(\ref{eq:ID_decomposed}) with respect to $V_G$:
\begin{equation}
  g_m = \underbrace{\frac{WC_{\rm ox}}{L}\,\bar{v}(\alpha)}_{\displaystyle g_m^{(n)}}
       + \underbrace{\frac{WC_{\rm ox}}{L}\,n\,
         \frac{d\bar{v}}{d\alpha}\,\frac{d\alpha}{dn}}_{\displaystyle g_m^{(\alpha)}},
  \label{eq:gm_decompose}
\end{equation}
where we used $dn/dV_G = C_{\rm ox}/q$ from Eq.~(\ref{eq:n_gate}).
The two terms have distinct physical origins:

\begin{itemize}
  \item $g_m^{(n)} = (WC_{\rm ox}/L)\,\bar{v}(\alpha)$: the conventional
    density-modulation term.
    At equilibrium ($\alpha = 0$), this equals the constant value
    $(WC_{\rm ox}/L)\,\bar{v}_{\rm eq}$, giving a flat $g_m$--$V_G$ curve
    in the linear regime.
  \item $g_m^{(\alpha)}$: a new distribution-shape-modulation term that
    arises because the gate voltage also controls $\alpha$ through $n(V_G)$.
    Its magnitude is proportional to
    $d\bar{v}/d\alpha = N_{\rm eq}N_{\rm neq}(\bar{v}_{\rm neq}-\bar{v}_{\rm eq})
    /(N_{\rm eq}+\alpha N_{\rm neq})^2$, which is positive whenever
    $\bar{v}_{\rm neq} > \bar{v}_{\rm eq}$.
\end{itemize}

The gate-voltage dependence of $g_m^{(\alpha)}$ is governed by
\begin{equation}
  \frac{d\alpha}{dn} = \frac{\alpha_{\rm max}\,h(V_D)\,n_c}{(n+n_c)^2},
  \label{eq:dalpha_dn}
\end{equation}
where $h(V_D) = V_D^2/(V_D^2+V_c^2)$.
To find the peak of $g_m^{(\alpha)}$ analytically, we substitute the
$\alpha(n)$ dependence into the expression for $d\bar{v}/d\alpha$ and
simplify.
Defining $\beta = \alpha_{\rm max}\,h(V_D)$, one finds
\begin{align}
  g_m^{(\alpha)} &\propto
  \frac{n}{\bigl(N_{\rm eq} + \beta N_{\rm neq}\,\tfrac{n}{n+n_c}\bigr)^{\!2}(n+n_c)^2}
  \nonumber \\
  &= \frac{n}{(A\,n + B)^2},
  \label{eq:gm_alpha_simplified}
\end{align}
where $A = N_{\rm eq} + \beta N_{\rm neq}$ and $B = N_{\rm eq}\,n_c$.
Setting $d/dn[n/(An+B)^2] = 0$ gives the exact peak density
\begin{equation}
  n_{\rm pk} = \frac{B}{A}
    = \frac{N_{\rm eq}\,n_c}{N_{\rm eq} + \alpha_{\rm max}\,h(V_D)\,N_{\rm neq}},
  \label{eq:npk}
\end{equation}
and the corresponding peak gate voltage
\begin{equation}
  V_G^{\rm pk} = V_{\rm th} + \frac{q\,n_{\rm pk}}{C_{\rm ox}}.
  \label{eq:VGpk}
\end{equation}
In the limit $\alpha_{\rm max} \to 0$, Eq.~(\ref{eq:npk}) gives
$n_{\rm pk} \to n_c$ and $V_G^{\rm pk} \to V_{\rm th} + qn_c/C_{\rm ox} \equiv V_G^*$.
For finite $\alpha_{\rm max}$, the peak shifts to lower gate voltages by the
factor $N_{\rm eq}/(N_{\rm eq} + \alpha_{\rm max}h(V_D)N_{\rm neq})$.
Since $N_{\rm eq} = k_BT$ and $N_{\rm neq} \approx \sigma\sqrt{\pi}$,
this shift encodes the hot-carrier energy width $\sigma$.
Consequently, $g_m^{(\alpha)}$ exhibits a peak at $V_G^{\rm pk}$---an anomalous
feature absent in equilibrium transport---and its position jointly reports
$n_c$ (the density crossover) and $\sigma$ (the hot-carrier spectral width).

\begin{figure*}[t]
  \centering
  \includegraphics[width=\textwidth]{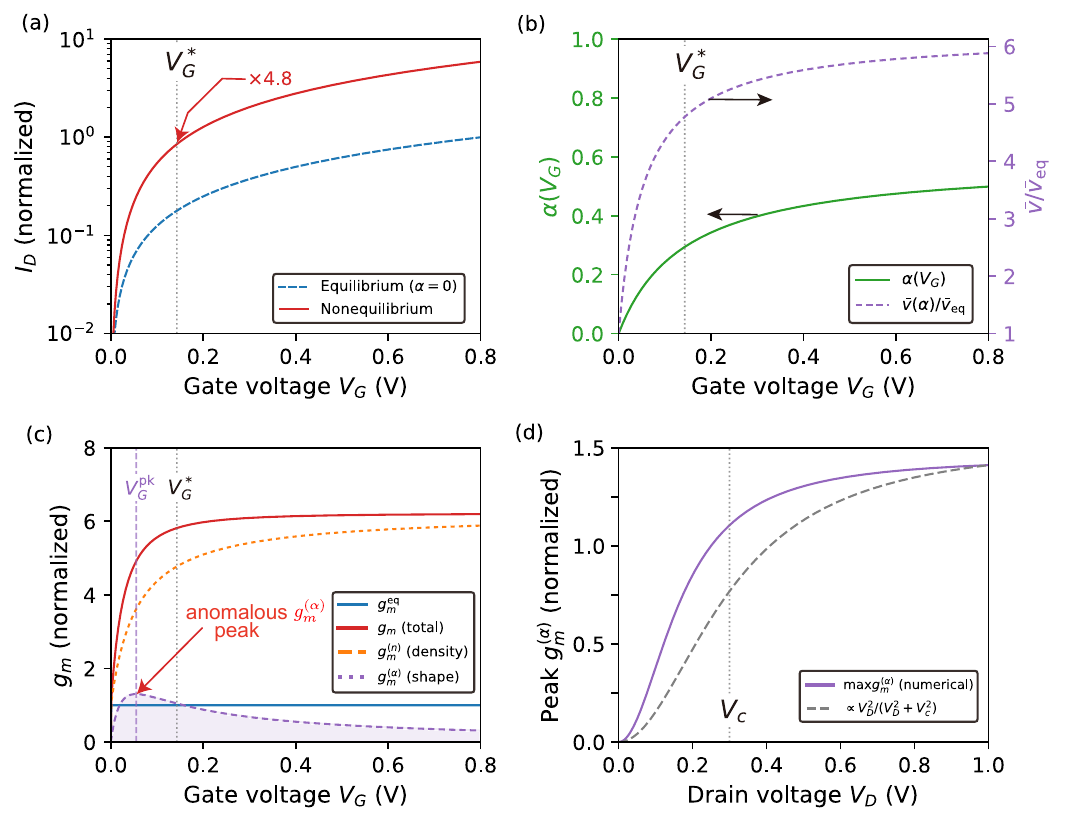}
  \caption{%
    Main steady-state prediction, combining the current--voltage and
    transconductance characteristics.
    (a) Drain current $I_D$ vs gate voltage $V_G$ at $V_D = 0.5$~V
    (logarithmic scale), normalized to the equilibrium saturation current.
    The nonequilibrium current (red solid) exceeds the equilibrium
    baseline (blue dashed) due to the hot-carrier velocity enhancement
    $\bar{v}(\alpha)$; the vertical dotted line marks $V_G^*$ where $n = n_c$.
    (b) Left axis: nonequilibrium amplitude $\alpha(V_G)$, showing saturation
    for $V_G \gg V_G^*$. Right axis: effective velocity ratio
    $\bar{v}(\alpha)/\bar{v}_{\rm eq}$, which reaches its maximum as
    $\alpha \to \alpha_{\rm max}h(V_D)$.
    (c) Transconductance $g_m(V_G)$ for the equilibrium ($g_m^{\rm eq}$,
    blue solid) and nonequilibrium (total, red solid) cases, decomposed
    into the density-driven term $g_m^{(n)}$ (orange dashed) and the
    distribution-shape-driven term $g_m^{(\alpha)}$ (purple dotted, filled).
    The equilibrium $g_m$ is flat (conventional) and the total nonequilibrium
    $g_m$ rises monotonically; the anomalous peak near $V_G^{\rm pk}$ resides in
    the shape contribution $g_m^{(\alpha)}$ (purple), isolated from the smooth
    monotone density background $g_m^{(n)}$.
    (d) Peak height of $g_m^{(\alpha)}$ vs drain voltage $V_D$, following
    $h(V_D) = V_D^2/(V_D^2+V_c^2)$ (gray dashed); the vertical dotted line
    marks $V_c = 0.3$~V.
  }
  \label{fig:steadystate}
\end{figure*}

\section{Results}
\label{sec:results}

We evaluate Eqs.~(\ref{eq:ID_decomposed})--(\ref{eq:gm_decompose}) numerically
for parameters representative of a monolayer MoS$_2$-like 2D semiconductor
($m^* = 0.5\,m_0$, spin--valley degeneracy $g = 4$, $T = 300$~K).
The gate dielectric is taken as 30~nm HfO$_2$ ($\varepsilon_r \approx 25$),
giving $C_{\rm ox}/q = 1.4\times10^{13}$~cm$^{-2}$V$^{-1}$.
The transport-kernel onset parameters are $E_b = 50$~meV and $\Delta = 20$~meV.
Hot carriers are characterized by $E_0 = 250$~meV, $\sigma = 40$~meV,
$\alpha_{\rm max} = 0.80$, $n_c = 2.0\times10^{12}$~cm$^{-2}$,
and $V_c = 0.30$~V.
Without loss of generality we set $V_{\rm th} = 0$, so that all gate voltages
are measured relative to threshold.
A finite $V_{\rm th}$ merely translates the $V_G$ axis and leaves every result
unchanged.

\subsection{Carrier distributions at fixed density}

Figure~\ref{fig:distributions}(a) compares $f_{\rm eq}(E)$ and $f_{\rm neq}(E)$
at the same carrier density $n$.
The equilibrium distribution decays exponentially with thermal energy $k_BT = 26$~meV,
while the nonequilibrium distribution is depressed at low energies
(due to normalization) but acquires a prominent Gaussian bump at $E_0 = 250$~meV.

Figure~\ref{fig:distributions}(b) shows the spectral current integrand
$j(E)f(E)$, which determines the contribution to $I_D$ at each energy.
The equilibrium integrand is concentrated at low energies where most carriers
reside, but the barrier $\mathcal{T}(E)$ strongly suppresses the contribution
below $E_b$.
The nonequilibrium integrand acquires a large additional peak at $E_0$,
because those hot carriers have both higher velocity $v(E_0) = v_T\sqrt{E_0/k_BT}$
and unit transmission $\mathcal{T}(E_0)\approx 1$.
The ratio of integrated spectral currents is $\bar{v}_{\rm neq}/\bar{v}_{\rm eq}
\approx 9.4$, meaning each hot carrier contributes $9.4\times$ more current
than a thermal carrier at $T = 300$~K.
This value 9.4 quantifies the per-carrier current contribution in the
\emph{pure} Gaussian component (i.e., $\bar{v}_{\rm neq}/\bar{v}_{\rm eq}$),
whereas the actual mixed-distribution current enhancement at
$\alpha = \alpha(n_0, V_D{=}0.5\,\text{V})$, weighted by the relative
populations of the two components, is the smaller value
$I_D^{\rm neq}/I_D^{\rm eq} \approx 5.1$ displayed in
Fig.~\ref{fig:distributions}(b).

\subsection{Steady-state transport prediction}

Figure~\ref{fig:steadystate}(a) shows $I_D(V_G)$ at $V_D = 0.5$~V.
Both curves increase linearly with $V_G$, as expected from $I_D \propto n$,
but the nonequilibrium current is enhanced by a factor that grows with
$V_G$ up to $n \sim n_c$ and then saturates.
This saturation is traced in Fig.~\ref{fig:steadystate}(b): as $n \gg n_c$,
$\alpha(n)$ saturates at $\alpha_{\rm max}h(V_D)$, and $\bar{v}(\alpha)$
approaches a constant.
The vertical line at $V_G^*$ marks the crossover density $n = n_c$.
At this gate voltage, the current enhancement is maximal in a differential
sense---the nonequilibrium $I_D$ increases faster with $V_G$ than the
equilibrium baseline, a hallmark of the $g_m^{(\alpha)}$ contribution.

The central experimental prediction is shown in Fig.~\ref{fig:steadystate}(c),
which plots $g_m(V_G)$ together with its decomposition.
The equilibrium transconductance is constant at $(WC_{\rm ox}/L)\,\bar{v}_{\rm eq}$
above threshold, as expected from conventional theory.
For the representative parameters used here, the total nonequilibrium $g_m$
itself rises monotonically with $V_G$ and
saturates at a higher plateau ($(WC_{\rm ox}/L)\,\bar{v}(\alpha_{\rm max}h(V_D))$)
than the equilibrium value, because the density term $g_m^{(n)}$ is monotone by
construction and dominates the magnitude.
The non-thermal signature is therefore carried not by the total $g_m$ but by its
\emph{shape-driven contribution} $g_m^{(\alpha)} = g_m - g_m^{(n)}$, which
exhibits a pronounced \emph{anomalous peak} at $V_G^{\rm pk}$
[Eq.~(\ref{eq:VGpk})] and vanishes in any density-only description (where
$g_m^{(\alpha)}\equiv 0$).
Because the density background $g_m^{(n)} = (WC_{\rm ox}/L)\,\bar{v}(\alpha)$ is
smooth and monotone, this localized peak appears as a shoulder on $g_m(V_G)$ and
is isolated by subtracting the calibrated monotone background.
It is this shape term---the gate-voltage derivative of the distribution
shape---that directly encodes spectral information beyond $n$.
This is the key signature of transconductance spectroscopy.
With the present parameters ($T = 300$~K, $\alpha_{\rm max} = 0.80$,
$\sigma = 40$~meV, $V_D = 0.5$~V), Eq.~(\ref{eq:npk}) gives
$n_{\rm pk}/n_c \approx 0.38$, so the peak appears at
$V_G^{\rm pk} \approx 55$~mV, well below the reference scale
$V_G^* = qn_c/C_{\rm ox} \approx 143$~mV.

The same panel decomposes $g_m$ into its two components.
The density-driven term $g_m^{(n)}$ is a smoothly increasing function that
inherits the $V_G$ dependence of $\bar{v}(\alpha)$, which rises monotonically
as $\alpha$ increases.
The shape-driven term $g_m^{(\alpha)}$ is localized near $V_G^*$ and
vanishes both below threshold ($n \to 0$) and for $n \gg n_c$
($d\alpha/dn \to 0$).
The anomalous peak thus resides entirely in $g_m^{(\alpha)}$, while the total
$g_m$ remains monotone---underscoring that the signature is a feature of the
distribution-shape contribution, isolated from the smooth density background,
rather than a maximum of the raw transconductance.

Figure~\ref{fig:steadystate}(d) shows the peak height of $g_m^{(\alpha)}$ as a
function of $V_D$.
It follows $h(V_D) = V_D^2/(V_D^2+V_c^2)$ (gray dashed), confirming
Eq.~(\ref{eq:dalpha_dn}).
For $V_D \lesssim V_c$, $g_m^{(\alpha)}$ grows quadratically with $V_D$,
while for $V_D \gg V_c$ it saturates.
This $V_D$-tunability provides an additional experimental signature:
measuring $g_m(V_G)$ at several $V_D$ values and fitting the peak height
to $h(V_D)$ extracts $V_c$, which encodes the electric-field scale for
hot-carrier generation.

\begin{figure*}[t]
  \centering
  \includegraphics[width=\textwidth]{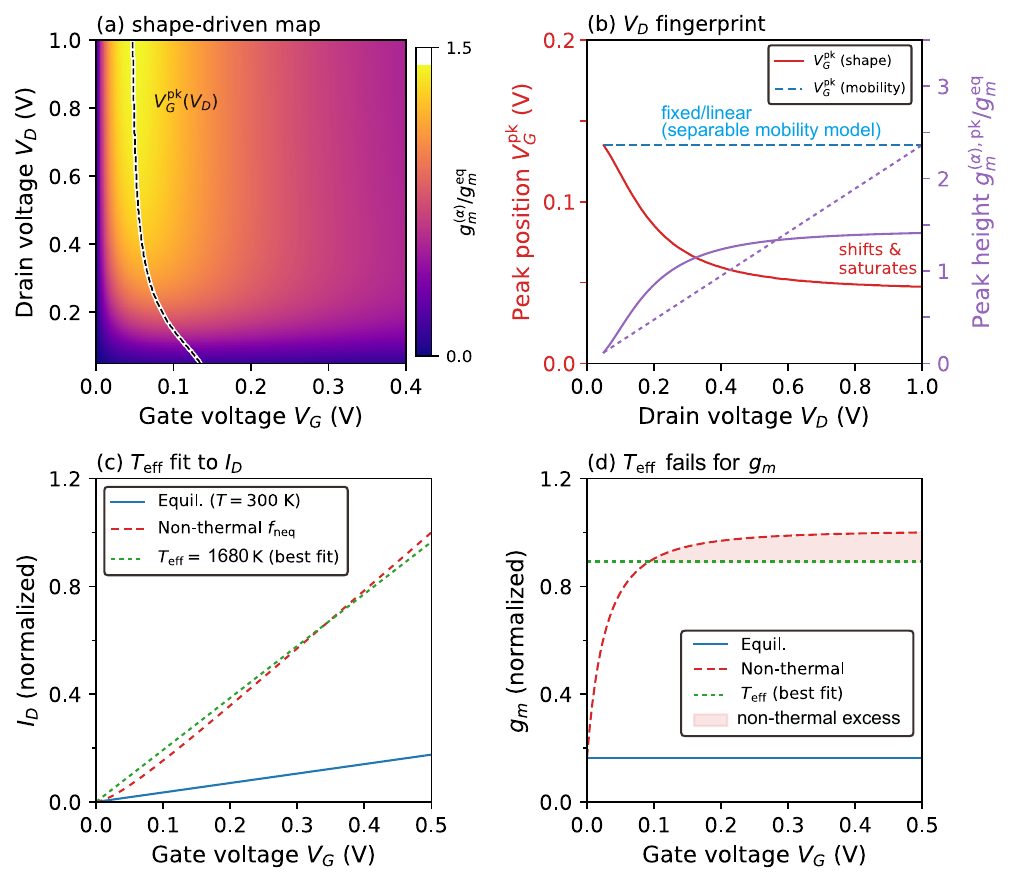}
  \caption{%
    Discrimination of the non-thermal distribution-shape peak from
    quasi-equilibrium descriptions.
    \emph{Top row---exclusion of a mobility-rolloff peak.}
    (a) Density map of the shape-driven transconductance
    $g_m^{(\alpha)}(V_G,V_D)$, normalized to the equilibrium plateau
    $g_m^{\rm eq}$; the overlaid curve is the peak trajectory
    $V_G^{\rm pk}(V_D)$, which shifts to lower $V_G$ ($\approx70$~mV from
    $V_D = 0.1$ to $1.0$~V) while the ridge intensity saturates at large $V_D$.
    (b) The two fingerprints versus $V_D$: the peak position $V_G^{\rm pk}$
    (red, left axis) shifts and saturates, and the peak height (purple, right
    axis) saturates as $h(V_D)$; a separable mobility model
    [Eq.~(\ref{eq:gm_mobility})] instead predicts a $V_D$-independent position
    (blue dashed) and a height linear in $V_D$ (purple dotted).
    \emph{Bottom row---exclusion of an effective-temperature model
    ($V_D = 0.5$~V).}
    (c) The best-fit heated-Boltzmann $T_{\rm eff}\approx1680$~K (green dash-dot)
    reproduces $I_D(V_G)$ reasonably well, but only by invoking an electronic
    temperature far above any quasi-equilibrium heating regime.
    (d) The same $T_{\rm eff}$ model gives a flat $g_m(V_G)$, whereas the
    non-thermal $g_m$ rises with $V_G$; the shaded region is the non-thermal
    excess over the flat $T_{\rm eff}$ prediction, unaccounted for by any
    effective-temperature model.
  }
  \label{fig:discrimination}
\end{figure*}

\subsection{Discrimination from quasi-equilibrium models}
\label{sec:discrimination}

The anomalous peak in the shape contribution $g_m^{(\alpha)}$ is the central
signature of the non-thermal mechanism, but it is not, by itself, unique to it.
Two quasi-equilibrium descriptions must be excluded before a peak extracted from
$g_m(V_G)$ can be attributed to the distribution shape: a gate-dependent
field-effect mobility $\mu(V_G)$, and an effective-temperature (heated-Boltzmann)
ensemble.
Both fail, and the way they fail provides the decisive, falsifiable
signature of the mechanism.

\paragraph{Exclusion of a mobility-rolloff peak.}
A gate-dependent mobility $\mu(V_G)$---from surface-roughness or phonon
scattering that strengthens with the vertical field~\cite{Kaasbjerg2012}---also
produces a $g_m$ peak. The two are separated by their drain-voltage dependence.
In the linear regime any quasi-equilibrium mobility description gives
$I_D = (qW/L)\,n\,\mu(V_G)\,V_D$, so
\begin{equation}
  g_m(V_G,V_D) = V_D\,\frac{qW}{L}\,\frac{d[n\,\mu(V_G)]}{dV_G}
  \equiv V_D\,G(V_G),
  \label{eq:gm_mobility}
\end{equation}
which is \emph{separable} in $V_G$ and $V_D$: the peak position is
$V_D$-independent and the peak height scales linearly with $V_D$.
The shape mechanism violates both. By Eqs.~(\ref{eq:npk})--(\ref{eq:VGpk}) the
peak position shifts to lower gate voltage as $V_D$ grows, while the peak height
saturates as $h(V_D)=V_D^2/(V_D^2+V_c^2)$ [Fig.~\ref{fig:steadystate}(d)].
For the present parameters the position shifts by
$\Delta V_G^{\rm pk}\approx 70$~mV between $V_D = 0.1$ and $1.0$~V
[Fig.~\ref{fig:discrimination}(a)], and the height saturates above
$V_D\sim V_c$ [Fig.~\ref{fig:discrimination}(b)].
The \emph{simultaneous} position shift and height saturation are not reproduced
by any separable, quasi-equilibrium mobility-rolloff model
[Eq.~(\ref{eq:gm_mobility})].
We note that a contrived contact or mobility response with strong,
nonseparable $V_D$ dependence could in principle mimic one of the two
signatures, so the discriminating power rests on their joint appearance rather
than on either alone.

\paragraph{Exclusion of an effective-temperature model.}
Previous treatments of hot carriers in FETs typically introduce an
effective electron temperature $T_{\rm eff} > T$ and replace $\mu$ by
$\mu(T_{\rm eff})$~\cite{Sze2007,Lundstrom2000}.
Such models assume that the heated distribution remains a displaced
Maxwellian, and therefore cannot capture the Gaussian hot-carrier component
in Eq.~(\ref{eq:fneq}), which represents a localized nonthermal excess
generated, e.g., by impact ionization, optical pumping at a specific photon
energy, or injection over a discrete barrier.
We find the optimal $T_{\rm eff}$ by minimizing the least-squares residual
in $I_D(V_G)$.
For the present MoS$_2$-like parameters at $V_D = 0.5$~V
this gives $T_{\rm eff} \approx 1680$~K, far beyond the regime in which a
simple quasi-equilibrium heating picture is physically transparent.
Yet even at this best-fit $T_{\rm eff}$, which reproduces $I_D(V_G)$ reasonably
well [Fig.~\ref{fig:discrimination}(c)], the model fails for
$g_m$ [Fig.~\ref{fig:discrimination}(d)]: because
$g_m^{T_{\rm eff}} = (WC_{\rm ox}/L)\,\bar{v}(T_{\rm eff})$
is a constant in $V_G$, the model predicts a flat transconductance, whereas the
non-thermal $g_m$ rises with $V_G$ and carries the localized shape contribution
$g_m^{(\alpha)}$ that a heated Boltzmann distribution cannot generate.
The shaded region in Fig.~\ref{fig:discrimination}(d) is the resulting non-thermal
excess of $g_m$ over the flat $T_{\rm eff}$ prediction---a direct
fingerprint of the non-thermal character of $f_{\rm neq}$ that no
effective-temperature description can reproduce.

\paragraph{Feasibility.}
The anomaly is large and the discriminator is undemanding.
The shape-driven peak reaches $g_m^{(\alpha),\rm pk}\sim(0.4$--$1.4)\,g_m^{\rm eq}$
for $V_D = 0.1$--$1.0$~V---an order-unity modulation of an already routinely
measured quantity, far above the $\lesssim1\%$ resolution of standard DC/lock-in
transconductance.
The diagnostic position shift of several tens of mV is resolved by sub-mV gate
steps, and because $E_0\gg k_BT$ the signature survives to room temperature.
The decisive measurement is thus a set of $g_m(V_G)$ sweeps at several $V_D$:
a peak that shifts and saturates identifies the shape mechanism, whereas a fixed
peak with $V_D$-linear height identifies conventional mobility rolloff.

\subsection{Two-time-scale separation}

Beyond the steady state, the framework predicts a specific transient
response to pulsed excitation.
In the simplest case a pump injects hot carriers at $t = 0$ and only the
hot-carrier amplitude relaxes, $\alpha(t) = \alpha_0\,e^{-t/\tau_E}$, while
the carrier density stays fixed.
This single-time-scale pump--probe limit, which yields the energy relaxation
time $\tau_E$ directly from the decay of $g_m(t)$, is treated in
Appendix~\ref{app:timedomain}.
In a more general scenario---such as a pulsed gate voltage or optical pump
that transiently injects both extra carriers and hot-carrier excitations---
two distinct transients appear simultaneously:
\begin{equation}
  I_D(t) =
  \bigl[n_0 + \Delta n\,e^{-t/\tau_n}\bigr]\,
  \bar{v}\!\bigl(\alpha_0\,e^{-t/\tau_E}\bigr),
  \label{eq:ID_two_tau}
\end{equation}
where $\tau_n$ is the carrier-density relaxation time (set by the device
RC time or dielectric relaxation) and $\tau_E$ is the energy relaxation time
(set by carrier--phonon scattering).
For $\tau_n \ll \tau_E$, these two time scales are separated and can be
independently extracted from a single $I_D(t)$ transient.

\begin{figure*}[tbp]
  \centering
  \includegraphics[width=\textwidth]{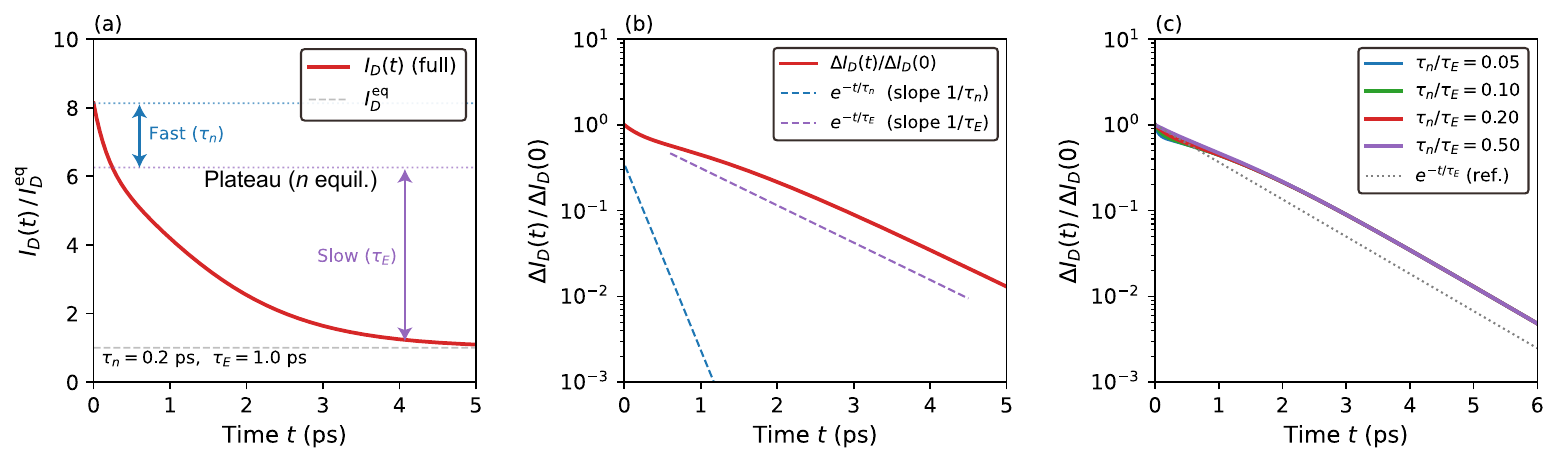}
  \caption{%
    Two-time-scale separation in the transient drain current after a
    pulsed excitation ($\Delta n/n_0 = 0.30$, $\alpha_0 = 0.60$).
    (a) $I_D(t)/I_D^{\rm eq}$ on a linear scale ($\tau_n = 0.2$~ps,
    $\tau_E = 1.0$~ps).
    The fast density decay ($\tau_n$) and slow spectral decay ($\tau_E$)
    are separated by an intermediate plateau where $n$ has equilibrated
    but hot carriers persist.
    (b) $\Delta I_D(t)/\Delta I_D(0)$ on a semi-logarithmic scale,
    revealing two linear segments with slopes $1/\tau_n$ (fast, blue dashed)
    and $1/\tau_E$ (slow, purple dashed).
    (c) Normalized anomaly for $\tau_n/\tau_E = 0.05$--$0.50$.
    All curves converge to the common $e^{-t/\tau_E}$ behavior (gray dotted)
    at long times, demonstrating clean extraction of $\tau_E$ when
    $\tau_n/\tau_E \lesssim 0.5$.
  }
  \label{fig:timescales}
\end{figure*}

Figure~\ref{fig:timescales}(a) shows $I_D(t)/I_D^{\rm eq}$ on a linear
scale for $\tau_n = 0.2$~ps and $\tau_E = 1.0$~ps.
Two regimes are apparent: a fast initial decay (duration $\sim\tau_n$) where
the excess density $\Delta n$ equilibrates, followed by a pronounced intermediate
plateau where the density has returned to $n_0$ but $\alpha$ is still large.
The current then decays slowly from the plateau to the equilibrium baseline
over the time scale $\tau_E$.
The height of the plateau encodes the amplitude $\alpha_0$ through
$I_D^{\rm plateau} = n_0\,\bar{v}(\alpha_0)$, providing an independent
determination of the initial hot-carrier injection level.

Figure~\ref{fig:timescales}(b) presents the same transient on a semi-logarithmic
scale, where the two exponential components appear as two linear segments
with distinct slopes $1/\tau_n$ and $1/\tau_E$.
This is the standard fingerprint of a bi-exponential decay, familiar from
optical pump--probe studies~\cite{Kash1985,Sun2008}, now accessible by
all-electrical measurements.
Fitting the long-time slope directly gives $\tau_E$ without contamination
from the fast density dynamics, provided measurements are taken at
$t \gtrsim 3\tau_n$.

Figure~\ref{fig:timescales}(c) shows the normalized anomaly
$\Delta I_D(t)/\Delta I_D(0)$ for a range of separation ratios
$\tau_n/\tau_E = 0.05$--$0.50$.
In all cases the long-time tail converges to a common $e^{-t/\tau_E}$
behavior (gray dotted reference), confirming that $\tau_E$ can be extracted
cleanly once $\tau_n/\tau_E \lesssim 0.5$.
In optimized high-bandwidth FET geometries the device-limited density
relaxation can in principle be made substantially shorter than the
carrier--phonon energy relaxation time $\tau_E \sim 1$--$5$~ps expected
for TMDs, placing this system within the resolvable regime.
The specific bandwidth requirements are discussed in Sec.~\ref{sec:discussion}.

\section{Discussion}
\label{sec:discussion}

\paragraph{Why transconductance spectroscopy is distinct from previous hot-carrier work.}
One might ask whether the present results are simply a restatement of
known hot-carrier physics in different notation.
They are not, for the following reasons.
First, Eq.~(\ref{eq:gm_spectral}) holds for \emph{any} $f(E)$, not just
the Boltzmann-plus-Gaussian model used here, provided the spectral
kernel $j(E)$ is gate-independent (corrections from a $V_G$-dependent
kernel are addressed below).
This means that $g_m(V_G)$ curves measured in any device, under any
excitation, are spectral transforms of the underlying distribution.
Second, the anomalous peak in $g_m^{(\alpha)}$ cannot be reproduced by any
density-only or quasi-equilibrium description without invoking an
ad hoc nonmonotonic $V_G$-dependence in $\mu(n)$ or
$T_{\rm eff}(n)$.
Indeed, a monotone $\mu(n)$ or monotone $T_{\rm eff}(n)$ necessarily
yields a monotone $g_m \propto \bar{v}(\alpha(n))\,(C_{\rm ox}/q)$ in the
linear regime.
By contrast, the distribution-shape mechanism predicts the peak
position, height, and $V_D$-dependence simultaneously from a single
unified parameter set $\{E_0,\sigma,n_c\}$, providing a tightly
constrained, falsifiable prediction.
Third, the measurement is all-electrical: it requires only a standard
lock-in $g_m$ sweep and is compatible with any gated 2D device,
with no need for pump lasers, time-resolved optics, or sub-kelvin cooling.

\paragraph{Robustness against gate-dependent kernels.}
The spectroscopic identity Eq.~(\ref{eq:gm_spectral}) was derived
assuming a gate-independent spectral kernel
$j(E) = v(E)\mathcal{T}(E)$.
In real devices, $V_G$ can also modulate $E_b$ (barrier lowering),
the contact transmission~\cite{Allain2015,Shen2021,Du2025}, or the band bending,
giving an additional kernel-modulation contribution
\begin{equation}
  g_m^{(j)} = \frac{qW}{L}\int_0^\infty
    f(E,V_G)\,\frac{\partial j(E,V_G)}{\partial V_G}\,dE,
  \label{eq:gm_j}
\end{equation}
so that in general $g_m = g_m^{(n)} + g_m^{(\alpha)} + g_m^{(j)}$.
For the Fermi barrier of Eq.~(\ref{eq:T}), $\partial j/\partial V_G =
-v(E)\,(\partial\mathcal{T}/\partial E)\,(\partial E_b/\partial V_G)$.
Since $\partial\mathcal{T}/\partial E$ is sharply peaked at $E = E_b$ (width
$\Delta$) with unit area, the kernel term reduces to
\begin{equation}
  g_m^{(j)} \simeq -\frac{qW}{L}\,\frac{\partial E_b}{\partial V_G}\,
    v(E_b)\,f(E_b;V_G).
  \label{eq:gm_j_est}
\end{equation}
Two consequences follow.
First, $g_m^{(j)}$ samples the distribution at the barrier edge $E_b$, not at
the hot-carrier energy $E_0 \gg E_b$.
Because the bump contributes
$\propto e^{-(E_b-E_0)^2/\sigma^2}\approx e^{-25}$ there, $g_m^{(j)}$ carries
\emph{no} peak structure from $f_{\rm neq}$ and adds only a smooth background to
$g_m(V_G)$---it cannot reproduce the anomalous peak.
Second, its magnitude is set by the barrier-lowering rate
$\partial E_b/\partial V_G$, an electrostatic/contact property that for a
well-gated channel in the on-state is small ($\partial E_b/\partial V_G \ll 1$).
The distinguishing signature of $g_m^{(\alpha)}$ from $g_m^{(j)}$ is
their qualitatively different $V_D$ dependence: $g_m^{(\alpha)}$ scales
as $h(V_D) = V_D^2/(V_D^2+V_c^2)$ [Fig.~\ref{fig:steadystate}(d)] and saturates
at large $V_D$, while $g_m^{(j)}$ contributions (barrier-lowering,
contact modulation) at fixed $V_G$ generally scale linearly or only
weakly in $V_D$.
Measuring the $V_D$-dependence of the anomalous peak thus provides an
in-situ separation of distribution-shape and kernel-modulation
mechanisms.
In practice, since contact and barrier-modulation effects can themselves
exhibit nonlinear $V_D$ behavior, the most robust criterion is not a
single $V_D$-dependence alone but the \emph{simultaneous} consistency of
peak position, peak height, and the joint fit to $I_D$ and $g_m$
(Appendix~\ref{app:simfit}, Fig.~\ref{fig:simultaneous})---the same
over-constrained signature that excludes the $T_{\rm eff}$ description.

\paragraph{Other device-related mechanisms and the over-constrained criterion.}
We emphasize that the appearance of a single peak in $g_m(V_G)$ is \emph{not},
by itself, claimed as a unique fingerprint of hot carriers.
Contact resistance and its bias dependence, interface traps, self-heating,
field- or density-dependent mobility degradation, and gate-induced barrier
modulation can each produce nonmonotonic features in the raw transconductance.
What discriminates the distribution-shape mechanism is not the peak alone but
the \emph{joint, over-constrained} behavior of several observables that these
parasitic effects do not reproduce together:
(i) the peak position $V_G^{\rm pk}$ shifts to lower gate voltage with
increasing $V_D$ [Eqs.~(\ref{eq:npk})--(\ref{eq:VGpk})], whereas a separable
mobility or trap response gives a $V_D$-independent position;
(ii) the peak height saturates as $h(V_D)$ rather than growing linearly in
$V_D$;
(iii) a single spectral parameter set $\{E_0,\sigma,\alpha_{\rm max},n_c\}$
must reproduce $I_D(V_G)$ and $g_m(V_G)$ simultaneously
(Appendix~\ref{app:simfit}); and
(iv) the low-bias limit $V_D \ll V_c$ must recover the flat, calibrated
equilibrium transconductance with $V_G^{\rm pk}\to V_G^*$.
Contact, trap, and self-heating artifacts are, moreover, suppressible or
separable by standard controls---four-probe or transfer-length contact
de-embedding, gate-bias- and frequency-dependent trap characterization, and
pulsed (low-duty-cycle) measurements that eliminate steady-state
self-heating.
The mechanism should therefore be regarded as identified by this diagnostic
protocol as a whole, not by any single peak.

\paragraph{Role of inelastic scattering and the transmission kernel.}
The energy-selective transmission $\mathcal{T}(E)$ in
Eq.~(\ref{eq:ID_general}) is an effective, elastic-like energy filter
calibrated from the equilibrium response; it does not resolve individual
inelastic collisions.
Inelastic scattering enters our description not through $\mathcal{T}(E)$ but
through the distribution itself---its steady-state hot-carrier amplitude
$\alpha$ and, in the transient case, its relaxation law
$\alpha(t) = \alpha_0 e^{-t/\tau_E}$.
The steady-state spectroscopy is therefore independent of the microscopic
collision integral: it reads out whatever nonthermal $f(E)$ is present,
however generated.
The time-domain extension (Sec.~\ref{sec:results};
Appendix~\ref{app:timedomain}) treats energy relaxation as a slow evolution of
this distribution parameter relative to the instantaneous transport response,
which is justified when the transit/RC time is short compared with $\tau_E$.
It is not a microscopic transport theory of energy-nonconserving events during
a single traversal, and it would break down if inelastic mean free paths became
comparable to the channel length, so that $\mathcal{T}$ itself acquired a
strong, energy-redistributing bias dependence. That regime lies beyond the
present kernel approximation.

\paragraph{Over-constrained simultaneous fit.}
Beyond excluding the effective-temperature description (Sec.~\ref{sec:discrimination}),
the non-thermal model provides an over-constrained demonstration: a
\emph{single} parameter set $\{E_0, \sigma, \alpha_{\rm max}, n_c\}$
simultaneously reproduces both $g_m(V_G)$ and $I_D(V_G)$, whereas the
$T_{\rm eff}$ model adds one free parameter yet already fails at the first
observable ($g_m$).
A quantitative simultaneous fit to synthetic data, confirming that the
spectral shape of $f(E)$---not its integrated density or effective
temperature---governs transport, is given in Appendix~\ref{app:simfit}
(Fig.~\ref{fig:simultaneous}).

\paragraph{Spectroscopic information from the $g_m$ anomaly.}
Equations~(\ref{eq:npk})--(\ref{eq:VGpk}) show that the peak position $V_G^{\rm pk}$
constrains two quantities:
(i) $n_c$, extracted from $V_G^{\rm pk}$ in the weak hot-carrier limit ($V_D \ll V_c$)
where $n_{\rm pk}\to n_c$; and
(ii) the Gaussian width $\sigma$ (via $N_{\rm neq}\approx\sigma\sqrt{\pi}$),
extracted from the shift $\Delta V_G^{\rm pk}(V_D) = V_G^* - V_G^{\rm pk}(V_D)$
as a function of $V_D$.
Independently, the peak height $g_m^{(\alpha),\rm pk} \propto J_{\rm neq}N_{\rm eq}
- J_{\rm eq}N_{\rm neq} \propto \bar{v}_{\rm neq} - \bar{v}_{\rm eq}$
is sensitive to $E_0$ through $v(E_0) = v_T\sqrt{E_0/k_BT}$, providing the
hot-carrier peak energy.
By performing $g_m(V_G)$ measurements at multiple $V_D$ values one can further
extract $V_c$, and temperature-dependent sweeps give $E_b$ and $\Delta$ from
$\bar{v}_{\rm eq}(T)$.
Taken together, a single set of $g_m$--$V_G$--$V_D$ measurements at varied $T$
constrains the parameters $\{n_c, \sigma, E_0, V_c, E_b, \Delta\}$, complementary
to optical techniques that typically access only one or two of these
simultaneously.
These extractions are model-assisted: they presuppose the calibrated
spectral-kernel form (Step~1) and a localized excess, so the inferred values are
best regarded as effective parameters of that description rather than uniquely
model-independent quantities. In particular, $E_0$ enters the peak height only
together with $\alpha_{\rm max}$ and the kernel, and is not fixed by the height
alone.

\paragraph{Two-time-scale separation as the third experimental pillar.}
Equation~(\ref{eq:ID_two_tau}) establishes
a hierarchy of electrical measurements for complete spectroscopy of $f(E)$.
The steady-state $g_m(V_G)$ sweep (Pillar~I: Fig.~\ref{fig:steadystate}) constrains the
spectral parameters $\{E_0, \sigma, n_c\}$ through the peak position and height.
The over-constrained consistency between $g_m$ and $I_D$ (Pillar~II)
excludes any effective-temperature description in the main text
[Fig.~\ref{fig:discrimination}(c,d)].
A quantitative simultaneous fit to synthetic data, which verifies that the
non-thermal model is strongly constrained by the joint $g_m$--$I_D$ data, is
provided as a supplementary demonstration in
Appendix~\ref{app:simfit} (Fig.~\ref{fig:simultaneous}).
The two-time-scale transient (Pillar~III: Fig.~\ref{fig:timescales}) then
isolates the energy relaxation time $\tau_E$ from the density relaxation time
$\tau_n$ in a single $I_D(t)$ measurement by exploiting the bi-exponential
structure of Eq.~(\ref{eq:ID_two_tau}).
Together these three observables constrain the full parameter set
$\{E_0, \sigma, n_c, \tau_E\}$ within the calibrated model from electrical
measurements alone, without relying on optical spectroscopy for readout.
Whereas Pillars~I and II rely on standard DC/lock-in sweeps,
Pillar~III requires GHz-to-THz-bandwidth time-domain instrumentation and is
therefore best viewed as a specialized extension that complements the
steady-state observables rather than replacing them.

\paragraph{Experimental measurement protocol.}
We outline a self-contained protocol by which the spectroscopic parameters
can be extracted sequentially from standard electrical measurements.
The minimum experimental dataset for verifying the non-thermal mechanism
and extracting $\{n_c, \sigma, V_c, E_0, \alpha_{\rm max}\}$ is a set of
$g_m(V_G)$ traces at several $V_D$ values (Steps~2--4 below), which
requires only a standard DC/lock-in FET sweep.
The time-domain extension (Step~5) is optional and yields the additional
parameter $\tau_E$.

\emph{Step 1 --- equilibrium calibration.}
At low drain voltage $V_D \ll V_c$, measure $I_D(V_G)$ and $g_m(V_G)$
at several temperatures $T = 200$--$350$~K.
The transconductance is flat and equals $(WC_{\rm ox}/L)\,\bar{v}_{\rm eq}(T)$,
giving the equilibrium velocity $\bar{v}_{\rm eq}(T)$, which follows the
thermally averaged spectral kernel and increases with temperature.
Fitting this temperature dependence to Eqs.~(\ref{eq:Jeq})--(\ref{eq:vbar})
with $\alpha = 0$ extracts the transport-kernel onset parameters $E_b$ and
$\Delta$.

\emph{Step 2 --- hot-carrier signature.}
Increase $V_D$ to values comparable to or exceeding $V_c$.
Subtract the smooth equilibrium-calibrated density background $g_m^{(n)}$ from
the measured $g_m(V_G)$.
If the residual shape contribution $g_m^{(\alpha)}$ develops a localized peak,
the effective-temperature description has broken down and the non-thermal model
is required.
The absence of such a peak at the same $V_D$ and $T$ would instead confirm
quasi-equilibrium conditions, providing a built-in null test.
Repeat at several $V_D$ values ($V_D = 0.2$--$0.7$~V recommended)
to track the peak height as a function of $V_D$.

\emph{Step 3 --- $V_D$-dependent peak height.}
Plot the peak height $g_m^{(\alpha),\rm pk}(V_D)$ against $V_D$.
A fit to $h(V_D) = V_D^2/(V_D^2 + V_c^2)$ [Eq.~(\ref{eq:dalpha_dn})]
extracts $V_c$, the electric-field crossover for hot-carrier generation.
The peak position $V_G^{\rm pk}(V_D)$ simultaneously yields
$n_{\rm pk}(V_D)$ via Eq.~(\ref{eq:n_gate}); fitting to
Eq.~(\ref{eq:npk}) with $V_c$ known then extracts $n_c$
and the spectral width $\sigma$ jointly.

\emph{Step 4 --- simultaneous fit and falsification.}
Using the parameters $\{E_b, \Delta, V_c, n_c, \sigma\}$ from Steps 1--3,
compute $I_D(V_G)$ and $g_m(V_G)$ at all measured $V_D$ values.
The non-thermal model should reproduce both observables with strongly
reduced parameter freedom: only $E_0$ and $\alpha_{\rm max}$ remain as
global fit parameters, both already constrained by Step 2.
As a falsification test, confirm that the best-fit $T_{\rm eff}$ model gives a
flat $g_m(V_G)$ and fails to capture the shape contribution $g_m^{(\alpha)}$
[Fig.~\ref{fig:discrimination}(d)].

\emph{Step 5 --- time-domain extraction of $\tau_E$.}
Apply a fast electrical pulse (gate or drain) with a sub-nanosecond to
picosecond-scale rise time and record $I_D(t)$ with high-bandwidth
(GHz-to-THz) instrumentation.
Fit the bi-exponential transient [Eq.~(\ref{eq:ID_two_tau})] to extract
$\tau_n$ (the fast, density-relaxation component, set by device RC) and
$\tau_E$ (the slow, spectral-relaxation component, set by carrier--phonon scattering).
The full parameter set $\{E_0, \sigma, n_c, V_c, E_b, \Delta, \tau_E\}$
is then constrained within the calibrated model from all-electrical
measurements.

\paragraph{Applicability and experimental scales.}
The method applies whenever a localized nonthermal excess is separated from the
transport onset by more than the thermal broadening,
$E_0 - E_b \gtrsim \text{few}\times k_BT$, so that the hot-carrier feature is
not washed out by the Boltzmann tail.
For the representative TMD parameters used here ($E_b = 50$~meV,
$E_0 = 250$~meV) this is comfortably satisfied at room temperature
($k_BT \approx 26$~meV), and the signature therefore survives to $300$~K and
above.
Lowering the temperature sharpens the spectral resolution but can enhance
competing trap and contact effects; raising it broadens both the thermal
background and the hot-carrier excess.
In gate voltage, the relevant density scale is $n \sim n_c \sim
10^{12}$~cm$^{-2}$, reached routinely in strongly gated, high-$\kappa$ TMD
FETs, and short channels ($L \sim 10$--$100$~nm) maximize the high-field region
that generates the excess.
The steady-state observables (Pillars~I and~II) require only standard
DC/lock-in instrumentation, whereas the optional time-domain extraction of
$\tau_E$ (Pillar~III) assumes energy-relaxation times $\tau_E \sim 0.1$--$10$~ps
typical of TMDs and hence GHz-to-THz measurement bandwidth.
Outside these ranges---strongly degenerate channels, $E_0$ comparable to $E_b$,
or dominant elastic disorder or spatial inhomogeneity (e.g.\ threshold
variation or contact nonuniformity) that smears the onset---the shape
contribution persists in principle but its peak may be too weak or too broad to
resolve.

\paragraph{Material and geometry considerations.}
The signal-to-noise ratio for detecting $g_m^{(\alpha)}$ is maximized when
$\bar{v}_{\rm neq}/\bar{v}_{\rm eq}$ is large, which requires $E_0 \gg E_b$
and $E_0 \gg k_BT$.
TMDs with high-$\kappa$ gate dielectrics (large $C_{\rm ox}$) shift $V_G^*$
to more accessible gate voltages.
Short-channel devices ($L \sim 10$--$50$~nm) increase the high-field region
and hence $\alpha$ at a given $V_D$.
For the pump--probe time-domain experiment, the bandwidth requirement is
$\sim1/\tau \sim 1$~THz for TMD materials~\cite{Kumar2014},
suggesting that THz near-field probing or GHz on-chip circuitry could
realize this measurement.

\section{Conclusion}
\label{sec:conclusion}

We have established that the transconductance $g_m$ of a 2D FET is
fundamentally a spectroscopic quantity.
It is the spectral current $j(E)$ weighted by the gate-voltage derivative of
the carrier distribution $f(E)$ and integrated over energy.
As such, it encodes the \emph{shape} of $f(E)$, not merely its integrated
density.
This spectroscopic identity [Eq.~(\ref{eq:gm_spectral})] is exact for a
gate-independent spectral kernel.
It gives rise to an anomalous peak in the distribution-shape contribution
$g_m^{(\alpha)}$, isolated from the smooth monotone density background of the
measured $g_m$.
This peak (i)~has no counterpart in equilibrium transport, (ii)~cannot be
reproduced by any quasi-equilibrium description without invoking ad hoc
nonmonotonic $V_G$-dependence in transport coefficients, and (iii)~constrains
the spectral parameters of the hot-carrier distribution through its position,
height, and drain-voltage dependence.

Two steady-state observables, accessible with standard DC/lock-in
instrumentation, provide the primary experimental verification of the
non-thermal character of the distribution.
Pillar~I is the anomalous peak in the shape contribution $g_m^{(\alpha)}$
(Fig.~\ref{fig:steadystate}).
Its decisive drain-voltage fingerprint---a peak that shifts to lower $V_G$
and saturates in height with $V_D$---distinguishes it from a conventional
mobility-rolloff peak, which would instead remain fixed in position and grow
linearly with $V_D$ [Fig.~\ref{fig:discrimination}(a,b)].
Pillar~II is the exclusion of any effective-temperature description, which
gives a flat $g_m(V_G)$ and fails to reproduce the shape contribution
$g_m^{(\alpha)}$ even at its best fit
[Fig.~\ref{fig:discrimination}(c,d)].
The over-constrained simultaneous fit of both $I_D$ and $g_m$ with a single
parameter set $\{E_0, \sigma, \alpha_{\rm max}, n_c\}$ provides a
supplementary quantitative confirmation
(Appendix~\ref{app:simfit}, Fig.~\ref{fig:simultaneous}).
Together these constrain the effective spectral parameters
$\{E_0, \sigma, n_c, V_c, E_b, \Delta\}$ without optical spectroscopic readout.
The energy relaxation time $\tau_E$ becomes accessible through a more
demanding time-domain extension---the bi-exponential transient $I_D(t)$
(Pillar~III, Fig.~\ref{fig:timescales})---which separates the energy
relaxation from the density relaxation $\tau_n$ but requires
GHz-to-THz-bandwidth instrumentation.

\textit{The transconductance provides a spectroscopic probe of carrier
distributions, revealing non-thermal features that lie beyond the reach
of any quasi-equilibrium description without ad hoc nonmonotonic
gate-voltage dependence.}
This capability positions gated 2D semiconductors as on-chip
distribution-function spectrometers, opening pathways to
(i)~systematic mapping of hot-carrier dynamics across phonon channels,
valley degeneracies, and temperatures in van der Waals monolayers;
(ii)~engineering of $g_m^{(\alpha)}$ peaks via dielectric and channel
geometry for energy-selective carrier amplification; and
(iii)~extending transconductance spectroscopy to multi-band and
spin-orbit-coupled 2D systems where distribution-shape effects couple
to valley and spin degrees of freedom.

\begin{acknowledgments}
The author thanks S. Adhikary and T. Kameda for useful discussions.
This work was supported by JSPS KAKENHI (Grants No.~JP25K01609,
No.~JP22H05473, and No.~JP21H01019). 
\end{acknowledgments}

\section*{Data availability}
The data that support the findings of this study, and the code used to
generate the figures, are available from the corresponding author upon
reasonable request.

\appendix

\section{Pump--probe transient at fixed carrier density}
\label{app:timedomain}

\begin{figure*}[tbp]
  \centering
  \includegraphics[width=\textwidth]{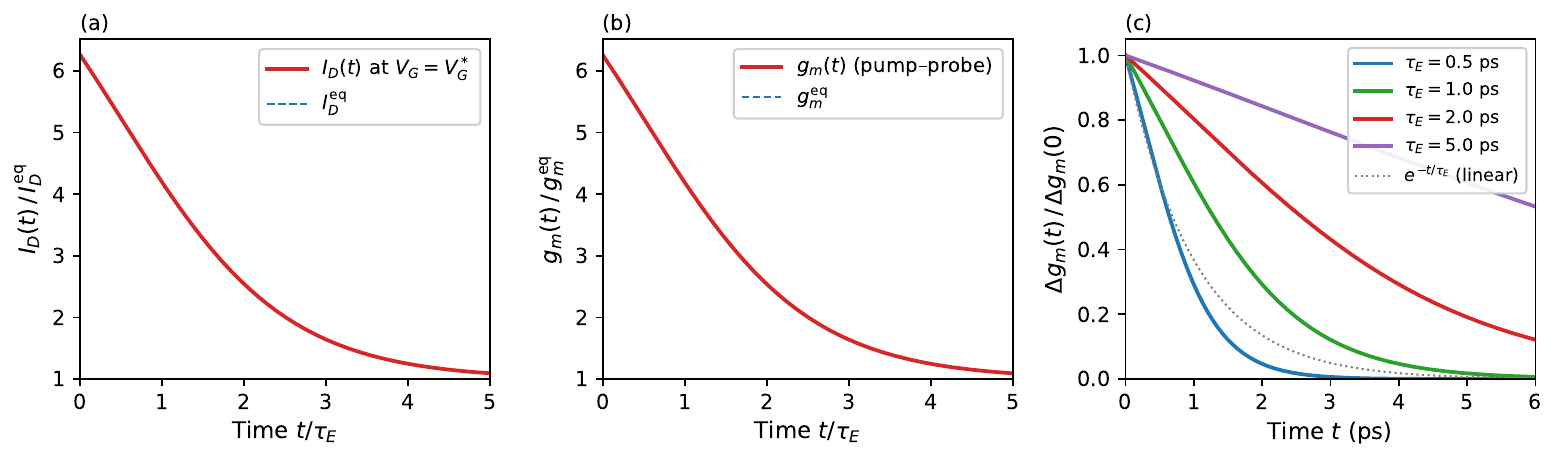}
  \caption{%
    Time-resolved response after pump excitation that injects hot carriers
    with $\alpha_0 = 0.6$ at $t = 0$, for $\tau_E = 1$~ps.
    (a) Drain current transient $I_D(t)/I_D^{\rm eq}$ at $V_G = V_G^*$.
    (b) Transconductance transient $g_m(t)/g_m^{\rm eq}$, which decays
    from the nonequilibrium value back to equilibrium as hot carriers relax.
    (c) Normalized transconductance anomaly $\Delta g_m(t)/\Delta g_m(0)$
    for relaxation times $\tau_E = 0.5$, $1.0$, $2.0$, and $5.0$~ps
    (blue, green, red, purple).
    The gray dotted line shows the pure exponential $e^{-t/\tau_E}$ for
    comparison, illustrating the slight nonlinearity due to $\bar{v}(\alpha)$.
  }
  \label{fig:timedomain}
\end{figure*}

\begin{figure*}[t]
  \centering
  \includegraphics[width=\textwidth]{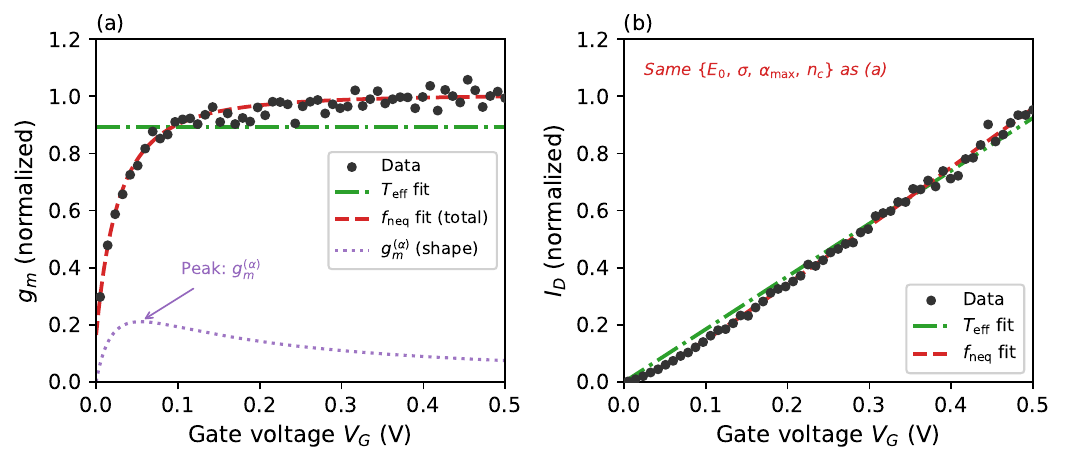}
  \caption{%
    Simultaneous fit of $g_m(V_G)$ and $I_D(V_G)$ using the non-thermal
    model $f_{\rm neq}$ (red dashed) versus the best-fit
    $T_{\rm eff} \approx 1680$~K model (green dash-dot).
    Black circles are synthetic data generated from the full non-thermal
    model with 3\,\% Gaussian noise added (realistic noise level for
    transport measurements).
    (a) $g_m(V_G)$: the non-thermal fit reproduces the data, including the
    shape contribution $g_m^{(\alpha)}$ (purple dotted) that carries the
    anomalous peak; the $T_{\rm eff}$ model gives a flat response with no such
    shape term.
    (b) $I_D(V_G)$: both models approximate the current, but only
    $f_{\rm neq}$ uses the \emph{same} four parameters as panel (a).
    The simultaneous fit over-constrains the model and demonstrates that
    spectral shape---not density or temperature alone---governs transport.
  }
  \label{fig:simultaneous}
\end{figure*}

\begin{figure*}[t]
  \centering
  \includegraphics[width=\textwidth]{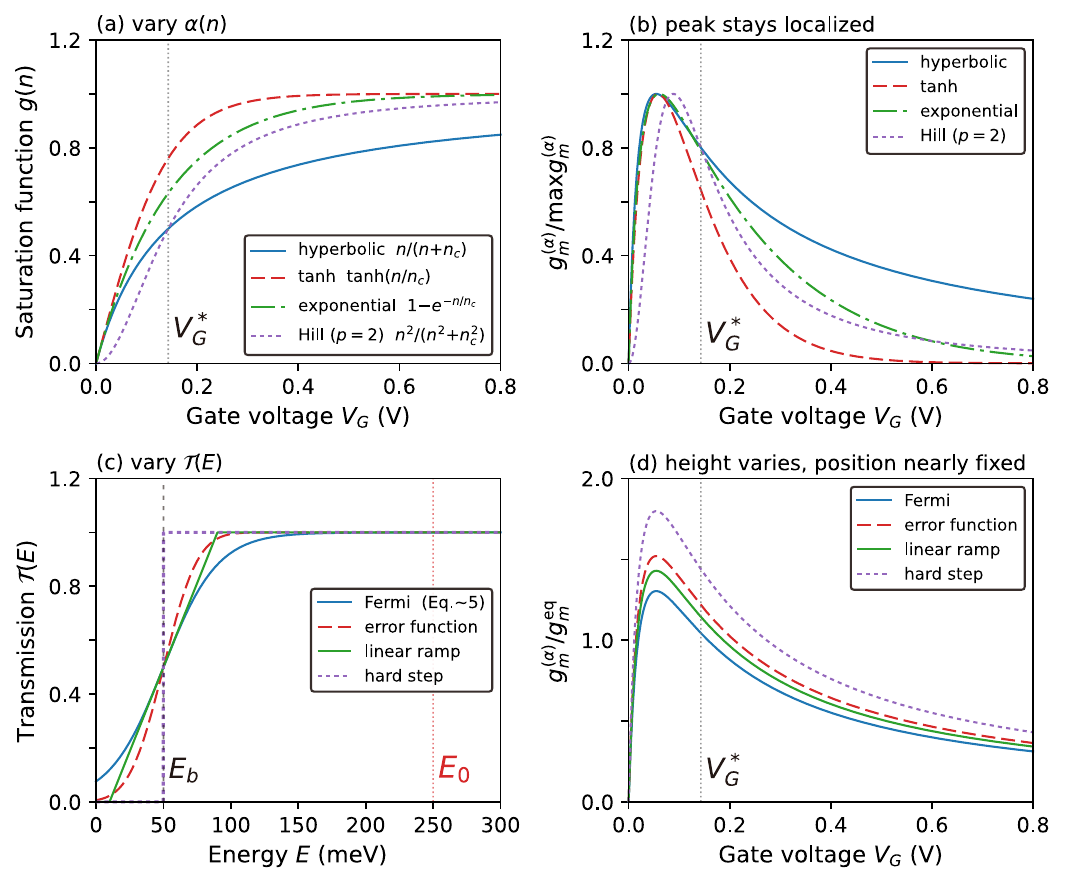}
  \caption{%
    Robustness of the $g_m^{(\alpha)}$ peak against the two phenomenological
    modeling choices.
    \emph{Top row---saturation function $\alpha(n)$.}
    (a) Four saturating forms of $g(n)$ versus gate voltage: hyperbolic
    $n/(n+n_c)$ (blue), $\tanh(n/n_c)$ (red dashed), $1-e^{-n/n_c}$
    (green dash-dot), and the Hill function $n^2/(n^2+n_c^2)$ (purple dotted).
    (b) Corresponding $g_m^{(\alpha)}(V_G)$, each normalized to its own peak:
    the peak \emph{position} stays near $V_G^*$ for every form.
    \emph{Bottom row---transmission $\mathcal{T}(E)$.}
    (c) Four transmission forms rising from $0$ to $1$ across $E_b$: Fermi step
    (blue), error-function (red dashed), linear ramp (green), and the
    discontinuous hard step $\Theta(E-E_b)$ (purple dotted); $E_0$ marks the
    hot-carrier energy.
    (d) Corresponding $g_m^{(\alpha)}(V_G)$, normalized to the equilibrium
    plateau $g_m^{\rm eq}$: the peak occurs at the \emph{same} gate voltage for
    all forms (position set by $\alpha(n)$, not $\mathcal{T}$); only the
    \emph{height} varies.
  }
  \label{fig:robustness}
\end{figure*}

The two-time-scale analysis of the main text reduces, in the limit of a
fixed carrier density, to a single-exponential pump--probe transient that
isolates the energy relaxation time $\tau_E$.
Consider a pump (optical or electrical) that injects hot carriers at $t = 0$,
establishing $\alpha(t{=}0) = \alpha_0$.
For $t > 0$ the hot carriers relax via carrier--phonon scattering,
$\alpha(t) = \alpha_0\,e^{-t/\tau_E}$, with $\tau_E$ the energy relaxation time.
In a pump--probe geometry, the gate is held fixed at $V_G = V_G^*$
and a small AC modulation $\delta V_G$ probes the instantaneous transconductance.
Because the pump-generated $\alpha$ is externally imposed and does not
depend on $V_G$, only the density-driven term survives:
\begin{equation}
  g_m(t) = \frac{WC_{\rm ox}}{L}\,\bar{v}\!\left(\alpha_0 e^{-t/\tau_E}\right).
  \label{eq:gm_t}
\end{equation}

Figures~\ref{fig:timedomain}(a) and (b) show the resulting transients in
$I_D(t)$ and $g_m(t)$ for $\tau_E = 1$~ps, normalized to their equilibrium values.
Both decay monotonically from their pump-excited values back to equilibrium.
The decay is exponential in $\alpha(t)$ but not strictly exponential in $I_D$
or $g_m$, because $\bar{v}(\alpha)$ is nonlinear in $\alpha$.
Fitting the transient to Eq.~(\ref{eq:gm_t}) therefore provides $\tau_E$
and simultaneously $\bar{v}_{\rm neq}/\bar{v}_{\rm eq}$, a ratio that
encodes $E_0$ (the hot-carrier energy) via Eqs.~(\ref{eq:Jeq})--(\ref{eq:vbar}).

Figure~\ref{fig:timedomain}(c) shows normalized transients
$\Delta g_m(t)/\Delta g_m(0)$ for several relaxation times $\tau_E$.
The curves are distinguishable over a decade in $\tau_E$ within a 6~ps window,
demonstrating that the measurement discriminates phonon-scattering channels
with sub-picosecond resolution, consistent with time-resolved Raman measurements
of longitudinal-optical (LO) phonon populations in III--V
semiconductors~\cite{Kash1985}.
Compared to optical pump--probe techniques~\cite{Sun2008,Pogna2016}
and recent electrical photoresponse measurements in van der Waals
heterostructures~\cite{Massicotte2016},
the present approach directly accesses the \emph{transport-weighted}
carrier distribution rather than the optical spectral weight or the
transient photocurrent, and can be combined with electrostatic gating to
tune $n$ and $V_G^*$ in situ.

\section{Simultaneous fit to synthetic data}
\label{app:simfit}

Figure~\ref{fig:simultaneous} demonstrates quantitatively that the
non-thermal model $f_{\rm neq}$ [Eq.~(\ref{eq:fneq})] with the \emph{single}
parameter set $\{E_0, \sigma, \alpha_{\rm max}, n_c\}$ simultaneously
reproduces both $g_m(V_G)$ [panel (a)] and $I_D(V_G)$ [panel (b)].
The two observables probe fundamentally different moments of $f(E)$: $g_m$ is
sensitive to $\partial f/\partial V_G$ and therefore to the \emph{shape} of
$f$, while $I_D$ tracks the total spectral weight.
A model that fits both with a single parameter set is therefore
over-constrained, providing unambiguous evidence that the spectral shape of
$f(E)$ governs transport, not merely its integrated density or effective
temperature.
The $T_{\rm eff}$ model cannot achieve this: it adds only one free parameter
yet already fails at the first observable ($g_m$).

\section{Robustness against modeling choices}
\label{app:alphaforms}

The two phenomenological ingredients of the model---the saturation function
$g(n)$ in the hot-carrier amplitude $\alpha(n)$ and the energy-selective
transmission $\mathcal{T}(E)$---are not derived from first principles.
We show here that the anomalous $g_m^{(\alpha)}$ peak is insensitive to both
choices.

\subsection{Saturation function $\alpha(n)$}

The phenomenological amplitude of Eq.~(\ref{eq:alpha}) uses the saturating
form $g(n) = n/(n+n_c)$.
The anomalous peak in the shape-driven transconductance does not rely on
this specific choice: any monotonically saturating $g(n)$ gives a
bell-shaped $d\alpha/dn$ and hence a localized $g_m^{(\alpha)}$ peak.
Figure~\ref{fig:robustness}(a,b) compares four representative forms---the
baseline hyperbola $n/(n+n_c)$, a hyperbolic tangent $\tanh(n/n_c)$, an
exponential saturation $1-e^{-n/n_c}$, and a Hill function
$n^2/(n^2+n_c^2)$---all normalized to rise from zero and saturate on the
scale $n\sim n_c$ [Fig.~\ref{fig:robustness}(a)].
The resulting $g_m^{(\alpha)}(V_G)$ curves [Fig.~\ref{fig:robustness}(b)], each
normalized to its own peak, all exhibit a localized peak near $V_G^*$ with
comparable width.
The forms with a steeper onset (tanh, Hill) place the peak slightly lower
in $V_G$ and decay faster at large $V_G$, but the qualitative
signature---a localized $g_m^{(\alpha)}$ peak driven purely by the distribution
shape---is common to all of them.
The peak \emph{position} is therefore a robust fingerprint of the saturating
hot-carrier generation mechanism rather than an artifact of the particular
analytic model adopted in the main text.

\subsection{Transmission $\mathcal{T}(E)$}

The main text models the transmission by the Fermi-function barrier of
Eq.~(\ref{eq:T}).
The anomaly does not rely on this form either.
The shape-driven term enters through the spectral current moments
$J_{\rm eq,neq} = \int j(E) f_{\rm eq,neq}(E)\,dE$ with
$j(E)=v(E)\mathcal{T}(E)$, and its sign and localization require only that the
hot-carrier component at $E_0 > E_b$ be transmitted more efficiently per
carrier than the thermal component, i.e.\ $\bar{v}_{\rm neq} > \bar{v}_{\rm eq}$.
Figure~\ref{fig:robustness}(c,d) compares four transmission forms that all
rise from $0$ to $1$ across $E_b$---the baseline Fermi step
$[1+e^{(E_b-E)/\Delta}]^{-1}$, an error-function step, a linear ramp, and the
discontinuous hard step $\Theta(E-E_b)$ [Fig.~\ref{fig:robustness}(c)].
The resulting $g_m^{(\alpha)}(V_G)$, normalized to the equilibrium plateau
$g_m^{\rm eq}$ of the same form [Fig.~\ref{fig:robustness}(d)], peaks at the
\emph{same} gate voltage for every transmission---the peak position is fixed by
$\alpha(n)$, not by $\mathcal{T}(E)$---while its height varies only mildly,
remaining of order the equilibrium response ($1.3$--$1.8\,g_m^{\rm eq}$) even
for the discontinuous hard step.
The transmission shape thus sets the magnitude of the anomaly but neither
its existence nor its location.
Taken together, the two robustness checks are complementary: the peak
\emph{position} is set by $\alpha(n)$ and is insensitive to $\mathcal{T}(E)$,
while the peak \emph{height} is set by $\mathcal{T}(E)$ and the hot-carrier
energy---so neither phenomenological input controls the existence of the
signature.

\section{Extension to linear (Dirac) dispersion}
\label{app:dirac}

The framework does not rely on the constant density of states of a parabolic
2D band.
For a linear (Dirac) dispersion $E = \hbar v_F |\mathbf{k}|$, the density of
states is $D(E) = g\,E/(2\pi\hbar^2 v_F^2) \propto |E|$ and the band velocity
is energy-independent, $v(E) = v_F$.
Writing the drain current in the same energy-resolved form
[Eq.~(\ref{eq:ID_general})], the transport now enters through the spectral
kernel
\begin{equation}
  j_{\rm D}(E) = D(E)\,v_F\,\mathcal{T}(E) \propto E\,\mathcal{T}(E),
  \label{eq:j_dirac}
\end{equation}
which replaces the parabolic-band kernel
$j(E) = v(E)\mathcal{T}(E) \propto \sqrt{E}\,\mathcal{T}(E)$.
In both cases $j(E)$ is a smooth, monotonically increasing energy filter that
favors high-energy carriers, so a localized nonthermal excess at $E_0 > E_b$ is
still weighted more heavily than the thermal population.
Consequently the transconductance retains the spectroscopic form
Eq.~(\ref{eq:gm_spectral}) and the decomposition
$g_m = g_m^{(n)} + g_m^{(\alpha)}$, and the anomalous $g_m^{(\alpha)}$ peak
persists.
Only the quantitative details change. Because the Dirac kernel grows linearly
rather than as $\sqrt{E}$, it weights the hot-carrier excess even more strongly,
so the per-carrier enhancement $\bar{v}_{\rm neq}/\bar{v}_{\rm eq}$, and hence
the peak height, is generally larger. The peak \emph{position}, set by
$\alpha(n)$ (Appendix~\ref{app:alphaforms}), is unchanged.
The same reasoning applies to any monotonic product $D(E)\,v(E)$, so the
mechanism is generic to gapped and gapless 2D materials alike.

\bibliography{refs}

\end{document}